\documentclass[12pt]{article}
\usepackage{graphicx}
\usepackage{amssymb}
\usepackage{epsfig}
\usepackage{lineno,hyperref}
\modulolinenumbers[5]
\usepackage{psfrag}
\usepackage{amsmath}
\usepackage{amssymb,subfigure}
\usepackage{xcolor}
\usepackage{color}

\definecolor{rred}{rgb}{0.7,0,0.1}
\definecolor{vgreen}{rgb}{0.2,0.6,0.3}

\newcommand{\sv}{\color{black}}
\newcommand{\sva}{\color{black}}
\newcommand{\svc}{\color{black}}

\begin{document}

\title{Statistical and Dynamical Properties of Covariant Lyapunov Vectors in a Coupled Atmosphere-Ocean Model - Multiscale Effects, Geometric Degeneracy, and Error Dynamics}
\author{St\'ephane Vannitsem$^{1}$ [\texttt{svn@meteo.be}]\\
$^1$ Institut Royal M\'et\'eorologique de Belgique, Brussels, Belgium\\
\\
Valerio Lucarini$^{2,3}$ [\texttt{valerio.lucarini@uni-hamburg.de}]\\
$^2$ Meteorologisches Institut, CEN, University of Hamburg,\\Hamburg, Germany\\
$^3$ Department of Mathematics and Statistics\\University of Reading, Reading, UK}
\date{\today}
\maketitle


\abstract
We study a simplified coupled atmosphere-ocean model using the formalism of covariant Lyapunov vectors (CLVs), which link physically-based directions of perturbations to growth/decay rates. The model is obtained via a severe truncation of quasi-geostrophic equations for the two fluids, and includes a simple yet physically meaningful representation of their dynamical/thermodynamical coupling. The model has 36 degrees of freedom, and the parameters are chosen so that a chaotic behaviour is observed. One finds two positive Lyapunov exponents (LEs), sixteen negative LEs, and eighteen near-zero LEs. The presence of many near-zero LEs results from the vast time-scale separation between the characteristic time scales of the two fluids, and leads to nontrivial error growth properties in the tangent space spanned by the corresponding CLVs, which are geometrically very degenerate. Such CLVs correspond to two different classes of ocean/atmosphere coupled modes. The tangent space spanned by the CLVs corresponding to the positive and negative LEs has, instead, a non-pathological behaviour, and one can construct robust large deviations laws for the finite time LEs, thus providing a universal model for assessing predictability on long to ultra-long scales along such directions. 
{\sva Interestingly,} the tangent space of the unstable manifold has 
{\sva substantial} projection on both atmospheric and oceanic components. 
{\sva The} results underline the difficulties in using hyperbolicity as a conceptual framework for multiscale chaotic dynamical systems, whereas the framework of partial hyperbolicity seems better suited, possibly indicating an alternative definition for the chaotic hypothesis. 
{\sva They also} suggest the need for accurate analysis of error dynamics on different time scales and domains and for a careful set-up of assimilation schemes when looking at coupled atmosphere-ocean models. 

\section{Introduction}

The climate provides a {\sv prominent} example of out-of-equilibrium, forced and dissipative complex system whose nonlinear dynamics is driven by the interplay of forcing, dissipation, and coupling across subdomains featuring sometimes vastly different physical and chemical properties and characteristic time scales of motion. The presence of instabilities and nonlinear interactions results in turbulent motions with variability spanning a very vast range of temporal and spatial scales. The instabilities are associated with transfers of energy across scales and between different reservoirs. Typically, exchanges of kinetic energy are associated with barotropic or shear instability, while transformation of available potential into kinetic energy is associated with baroclinic or to convective instability. The fact that organized motions results from the presence of thermal gradients is the key ingredient for interpreting the climate as a heat engine \cite{Lucarini2009,Lucarini2014}.

These features are associated with the presence of a limited range of deterministic predictability --  \textit{e.g.} \cite{Lorenz1982, Nicolis1995, Vannitsem1997, Tribbia2004} and references therein -- which provides fundamental constraints to our ability to, \textit{e.g.}, predict accurately the weather in the mid-latitudes with a lead time larger than, say, 10 days, except in exceptional circumstances, and poses formidable challenges in the tantalizing effort aimed at building more and more accurate models. The presence of multiple scales of motions clearly suggests that geophysical fluid dynamics often provides examples of stiff problems. The quest for achieving predictive skill requires, apart from a continuous computational brute force effort leading to reducing the spatial scales directly resolved by the model, the need for devising accurate  (and efficient) numerical schemes, and, very importantly, parametrizations describing the unresolved subgrid scale processes such as (scale-dependent) turbulent diffusion and mixing, phase changes, transport at interfaces (\textit{e.g.} between atmosphere and ocean or land surface), and radiation absorption, scattering, and emission. These issues are extremely relevant also in the context of comparing model outputs and observations and of assimilating observational data in models.

Methods of scale and asymptotic analysis typically try to derive from the most general set of evolution equations for geophysical fluids some simplified laws which are approximately valid within certain regimes and certain temporal and spatial scale \cite{Klein2010}. This leads to projecting out motions and related instabilities occurring outside the chosen dynamical range and losing information of the scale-scale interaction. Arguably, the most successful example of application of this approach has been the derivation of quasi-geostrophic dynamics, which {\sv was} at the base {\sv of the early developments } of the numerical weather forecast \cite{Lynch2008}.

At practical level, in many cases, one would like to be able to construct reduced order model able to describe the most important features of the large scale geophysical motions, and, find theoretically or empirically a simplified representation of the effect of smaller scale motions in the form of parametrizations \cite{Franzke2015}. Unfortunately, since climatic variability does feature a continuum spectrum with no gaps, common methods suited to performing mode reduction and aimed at findings efficient descriptions of the impact of small/fast scales on the large/slow scales cannot be readily applied, as the commonly adopted time scale separation hypothesis does not hold. Therefore, in the construction of parametrizations, in addition to stochasticity \cite{Majda2001,Palmer2009}, non-markovian effects need to be taken into account \cite{Wouters2012,Wouters2013,Chekroun2015a,Chekroun2015b}. The problem is extremely interesting and complex when one wants to construct a simplified representation of the coupling between subsystems having different characteristic time scales, like the atmosphere and the ocean. In this case, one aims at being able to have an approximate yet reasonable representation of the quasi-independent dynamics of the two subdomains, plus, critically, of the processes where the two cannot be easily separated.

An additional problem one faces in understanding the properties of geophysical flows is that classical approaches focus typically on studying linear stability of given stationary reference background flows and study the associated unstable processes and energetics supporting the growth of instability \cite{Pedlosky1987,Vallis2006}. Nonetheless, as mentioned above, the actual flows are typically turbulent, so that such a picture retains only a qualitative yet extremely instructive value. 
{\sva In} recent years, the theory of dynamical systems and statistical mechanics have provided tools able to bridge at least partially the gap between the presence of high-dimensional chaos in geophysical systems and the need for studying accurately and providing physical interpretation to the departure of the trajectory from a reference dynamically generated background aperiodic state.

The Covariant Lyapunov Vectors (CLVs), first introduced in \cite{Ruelle1979},  later discussed in \cite{Trevisan1998}, and practically made available thanks to recent proposed algorithms \cite{Ginelli2007,Wolfe2007,Kuptsov2012,Froyland2013}, are a norm-independent and covariant basis of the tangent linear space, providing a splitting between the unstable manifold, describing the unstable perturbations leading to the divergence of the trajectories, the neutral manifold, typically corresponding to the direction of the flow, and the stable manifold, which corresponds to the contracting directions. The CLVs provide also a natural basis for constructing the response operator describing how the invariant measure of a chaotic statistical mechanical system is affected by perturbations to its dynamics \cite{Ruelle2009}.

The CLVs are defined through a suitable geometric construction involving both the forward and the backward Lyapunov vectors, whose computation can be seen as a byproduct of the usual Benettin et al. algorithm \cite{Benettin80} used for estimating of Lyapunov Exponents (LEs). The size of the perturbations oriented according to the CLVs grows or decays with an approximate exponential law, where the averages of the fluctuating rates of growth or decay correspond one-to-one to the LEs. As opposed to the CLVs, the forward and backward Lyapunov vectors are not covariant, so that it is hard to interpret them physically \cite{Pazo2010}. Therefore, CLVs allow for associating a time-dependent field to each LE, thus providing a connection between observed rates of growth and decay of perturbations and the corresponding physical modes of the system. In the case of spatially extended systems, this also allows for associating time and spatial scales of perturbations, and investigating their localization properties \cite{Pazo2008}. 

Note that while the LEs are the same no matter whether one uses the covariant, forward, or backward construction of the corresponding vectors, the finite time LEs (FTLEs) in general do differ   \cite{Laffargue13,Politi15}. Using the CLV formalism, the possibility of linking instantaneous rates of growth/decay of the perturbations and spatial patterns makes it possible to provide a physical interpretation of specific conditions of the flow supporting anomalous properties in terms of predictability.  In particular, one might find insightful signatures of the conditions supporting, \textit{e.g.},  sustained negative anomalies of the value of the instantaneous estimate of a given LE. In this case, the fluid configuration might support enhanced predictability, and the corresponding CLV might have a special structure associated with anomalous transport or energy exchange mechanisms, see \cite{Schubert2015a, Schubert2015b}.

As CLVs are dynamically generated, they might provide a suitable basis for constructing reduced order models, with the unstable modes being the obvious candidates for explaining a large part of the variability of the system, and with the possibility of selectively removing certain scales of motions by not including the corresponding CLVs in the reduced model \cite{Schubert2015a}. Hence, one can think of using CLVs to test a posteriori the validity of scale and asymptotic approximations in a given regimes determined by, \textit{e.g.} a specific choice of non-dimensional numbers, and to construct parametrizations.

In the case of multiscale chaotic systems, following Gallavotti \cite{Gallavotti2014}, one expects to be able to separate different scales of motions by looking at the spectrum of LEs and at the properties of the corresponding CLVs. One should be able to typically associate small-scale instabilities to large LEs, and, conversely, large scale instabilities to small LEs, with the former ones localized in space and quickly decorrelating, and the latter ones corresponding to the dominant large-scale patterns of motion of the system.

In a previous paper,  the potential of CLVs analysis for providing a very thorough analysis of quasi-geostrophic atmospheric dynamics, able to provide new insights on  baroclinic and barotropic instabilities and the energetics of the atmospheric circulation \cite{Schubert2015a}, has ben demonstrated. 
In a more recent paper, CLVs have proved very useful for studying the dynamics and
predictability of blocking events and the associated energetics. 
{\sva The} CLVs can {\sva therefore}  add
geographic information to the analysis of the instabilities and provide hints for studying
coupling mechanisms and feedbacks \cite{Schubert2015b}.
Therefore, we can metaphorically construct the \textit{climate} (intended as the average properties) of the \textit{weather} (intended as fluctuations due to unstable processes or decaying modes). 
Unfortunately, quasi-geostrophic equations do not provide, by construction, a suitable environment for support multi-scale processes, as they result from a severe asymptotic expansion of the incompressible Navier-Stokes equations in a rotating frame of reference \cite{Pedlosky1987,Klein2010}. In fact, one can see as one of the results of \cite{Schubert2015a,Schubert2015b} exactly the a-posteriori confirmation of the effectiveness and self-consistency of the quasi-geostrophic approximation. 

In this paper we want to take up the challenge of addressing, instead, multi-scale effects in a very simple model providing a metaphor of the coupled atmosphere-ocean system \cite{Vannitsem2015}. The idea is to study the time scale separation between the two subsystems resulting from the diversity of the thermodynamic properties of the two geophysical fluids through the analysis of the LEs and of the corresponding CLVs. In particular, we would like to be able to associate fast scales motions, as described by large LEs, to CLVs projecting almost entirely (but not exclusively) on the atmospheric component of the system, and slow motions, associated with CLVs projecting predominantly on the oceanic component of the system, or, most interestingly, on both components. In the latter case, we would be able to define rigorously coupled modes of instability, beyond the usual linear feedback analysis. 

Additionally, we wish to investigate accurately the geometrical properties of the tangent space and assess  whether strong degeneracies exist between the directions defined by the CLVs , so that coupling between different modes can be easily activated as soon as very small yet finite perturbations are considered \cite{Takeuchi2011}. 

Such a geometrical analysis is complemented by the investigation of the properties of the FTLEs corresponding to the CLVs: we will look at whether significant time correlations exist between FTLEs corresponding to different CLVs, thus indicating the possibility of collective fluctuations of growth or decay rate of perturbations involving several modes. Additionally, following \cite{Pazo2013,Schalge2013}, we will test whether long but finite time averages of the FTLEs obey large deviations laws \cite{Kifer90,Touchette09}, which gives us a solid mathematical framework for assessing the fluctuations of predictability at different (long) time scales and, at a more basic level, to what extent hyperbolicity holds. At more practical level, we will study the related issue of the dynamics of error using the so-called $L^2$ and logarithmic norms, and emphasize how the CLV formalism helps us in the understanding of this aspect of the chaotic flow 
{\sva investigated here}.

{\sva This} exercise has relevance for addressing fundamental properties of multiscale chaotic dynamical systems and, specifically, for coupled atmosphere-ocean systems, and for the important task of constructing conceptual and practical tools for performing the so-called coupled data assimilation \cite{Sogiura2008}. In fact, the message is that in order to be able to predict accurately the state of the atmosphere (ocean), information on the state of the ocean (atmosphere) can be of extreme relevance, with as determined by the structure of the CLVs. Moreover, the analysis of the coupled atmosphere-ocean modes as determined by the mixed CLVs might prove crucial for addressing the challenges of seasonal prediction.

This paper is structured as follows. In Sec. \ref{sec:formulation} we recapitulate at heuristic level the main ingredients theory of CLVs and LEs. In Sec. \ref{ssec:full} we provide a description of the reduced order coupled atmosphere-ocean model used in this study. In Sec. \ref{sec:results} we present the results of  our investigation, describing the properties of the CLVs and of the corresponding LEs, the geometric structure of the tangent space,  the statistical properties of the FTLEs, with a specific emphasis of the derivation of large deviation laws for long time averages of the FTLEs. As it turns out, this analysis allows to identify three subspaces, one corresponding to the directions that (asymptotically) expand, one corresponding to the directions that (asymptotically) contract, plus a center direction involving several CLVs with LEs close to $0$, in which the dynamics is highly non-
trivial. The CLVs corresponding to the unstable manifold and the center direction have projections of both the atmospheric and oceanic components, thus showing the relevance of coupled processes regarding the predictability of the system on both short and long time scales. Section \ref{sec:error} describes the dynamics of the error along the vectors emerging in these subspaces. Finally in Sec. \ref{sec:concl} we present our conclusions and perspectives for future work, and we summarize the implications of the results  when dealing with multi-scale systems as the coupled ocean-atmosphere one.

\section{A Brief Introduction to Covariant Lyapunov Vectors}
\label{sec:formulation}
For the benefit of the reader and in order to introduce our notation, in this section we provide a brief and rather informal recapitulation of LEs and CLVs. Much more rigorous and complete treatments of these topics can be found in, \textit{e.g.}, \cite{Eckmann1985,Ginelli2007,Wolfe2007,Kuptsov2012,Froyland2013}. Let's consider a sufficiently well-behaved autonomous dynamical system \footnote{One might assume we are treating Axiom A systems [45] or studying high dimensional systems for which, according to the chaotic hypothesis, effective Axiom A properties can be assumed [14,15]. We will critically discuss these aspects later in the paper.} 

\begin{equation}
\frac{d\vec{x}}{dt} = {\vec F}(\vec x, \lambda)
\label{equat}
\end{equation}
where $\vec x$ is the set of variables $\vec x$ = $( x_1, ..., x_N)$ defining the phase space,
$\vec F$ stands for the evolution laws and $\lambda$ denotes a set of control parameters. 
We write the formal solution of Eq. (\ref{equat}) as
\begin{equation}
\vec x (t) = f_t (\vec x(t_0), \lambda) 
\label {traject}
\end{equation}
In the real world, the initial state $\vec x(t_0)$ cannot be specified with an infinite accuracy.
It is subjected to an initial error $\delta \vec x$ whose subsequent evolution is described by a linearized
system of equations provided that its initial magnitude is sufficently small,
\begin{equation}
\frac{d\delta \vec x}{dt} =  \frac{\partial \vec F}{\partial \vec x}_{\vert \vec x(t)}
\delta \vec x
\label {linear}
\end{equation}
The formal solution of this equation can be written as 
\begin{equation}
\delta \vec x (t) = {\bf M}(t,\vec x(t_0)) \delta \vec x (t_0)
\label{fund}
\end{equation}
where $\bf{M}$ is the fundamental matrix. 

By virtue of the Oseledec theorem \cite{Eckmann1985}, asymptotic quantities independent of $t$ or $t_0$ can 
be defined.  The theorem says that there exists a matrix $S_{\vec x(t_0)}$ such that

\begin{equation}
S_{\vec x(t_0)} = \lim_{t \rightarrow \infty} ({\bf M^T}(t,\vec x(t_0)) {\bf M}(t,\vec x(t_0)))^{1/2(t-t_0)}
\end{equation}
where ${\bf M}^T(t, \vec x(t_0))$ is the transpose of $\bf{M}${\sv, which depends on the specific scalar product used, {\svc and which is the fundamental matrix of the adjoint model}.
} 
The logarithm of the eigenvalues of the matrix $S$ are the Lyapunov exponents (LEs) $\sigma_j$ and  
the corresponding eigenvectors, the forward Lyapunov vectors, $l^+_i$. The corresponding LEs will be
referred to as the forward Lyapunov exponents (FLEs) when needed.

Similarly, there exists a matrix $S'_{\vec x(t)}$ such that
\begin{equation}
S'_{\vec x(t)} = \lim_{t_0 \rightarrow - \infty} ({\bf M}(t,\vec x(t_0)) {\bf M^T}(t,\vec x(t_0)))^{1/2(t-t_0)}
\end{equation}
whose eigenvalues are the same as the ones of $S_{\vec x(t_0)}$. The corresponding
eigenvectors are called the backward Lyapunov vectors, $l^-_i$. The corresponding LEs will be
referred to as the backward Lyapunov exponents (BLEs) when needed.
Forward and backward vectors are both local properties of the flow since they depend on $\vec x(t_0)$ and 
$\vec x(t)$, respectively. Assuming ergodicity, one has that the LEs are the same for almost all $\vec{x}{(t_0)}$, so that the space-dependence can be dropped. 

Matrices $S$ and $S'$ can be evaluated at the same place along the reference trajectory $\vec x(t')$ and
one can determine the orthogonal eigenvectors of these symmetric matrices, $\vec l^+_i (\vec x(t'))$ and
$\vec l^-_i (\vec x(t'))$. There exist subspaces $W_i (\vec x(t'))$ such that

\begin{equation}
W_i(\vec x(t')) = \vec l^-_1 \oplus ... \oplus \vec l^-_i \cap \vec l^+_i \oplus ... \oplus \vec l^+_N
\label{wi}
\end{equation}
where $\oplus$ is the direct product \cite{Ruelle1979}.

These new subspaces have the important properties that their evolution under the fundamental matrix is

\begin{equation}
{\bf M} (\tau ,\vec x(t')) W_i(\vec x(t')) = W_i (\vec x(\tau)) 
\end{equation}

When the Lyapunov spectrum is non degenerate, one can define an {\sv arbitrary} vector $\vec g_i$ {\sv within $W_i(\vec x(t'))$} such that

\begin{equation}
{\bf M} (\tau ,\vec x(t')) \vec g_i(\vec x(t')) = \alpha_i (\tau, \vec x(t')) \vec g_i (\vec x(\tau)) 
\label{ampli}
\end{equation}
where $\alpha_i (\tau, \vec x(t'))$ is the amplification factor. Note first that the basis $\{ \vec g_i\}$ do
not form an orthogonal basis and also that in the long time
limit, the amplifications give access to the LEs,

\begin{equation}
\sigma_i =  \lim_{(\tau-t') \rightarrow \infty} \frac{1}{\tau-t'} ln \,\, (\alpha_i (\tau, \vec x(t')))=
 \lim_{(\tau-t') \rightarrow \infty} \sigma_i^{\tau-t'}(\vec x(t'))) 
\end{equation}

where we indicate by $\sigma_i^{\tau-t'}(\vec x(t'))) $ the average of the growth rate taken over a time window $ \tau-t'$ starting at $t'$ at position $\vec x(t'))$, and the space-dependence is dropped in the limit. 

We refer to $\sigma_i^{\tau-t'}(\vec x(t'))) $ as the $i^{th}$ finite time Lyapunov exponent (FTLE). The vectors $\{ \vec g_i\}$, already referred to as CLVs,  characterizes the local stability along the reference trajectory and constitutes
therefore a basis to describe the evolution of small perturbations along the stable and unstable manifolds. Therefore, we refer to the corresponding LEs as covariant Lyapunov exponents (CLEs).

The evolution of any (small) perturbation, $\vec {\delta x} (t_0) = \sum_{i=1}^{N} c_i \vec g_i(\vec x(t_0))$, 
can {\sva then} be described as 

\begin{equation}
\delta \vec x(\tau) = \sum_{i=1}^{N} \alpha_i (\tau, \vec x(t_0)) c_i \vec g_i(\vec x(\tau)) 
\end{equation}

One can also define a local stretching factor $\chi_i (\vec x(t))$ such that,

\begin{equation}
\alpha_i (\tau, \vec x(t_0)) = \exp \int_{t_0}^{\tau} \chi_i (\vec x(t))  \mathrm{d}t
\end{equation}
where, clearly we have $$ \lim_{(\tau-t_0) \rightarrow 0}  \sigma_i^{\tau-t_0}(\vec x(t_0)))=\sigma_i^0(\vec x(t_0)))=\chi_i (\vec x(t_0)),$$ which defines the instantaneous $i^{th}$ FTLE.
Note that amplification factors, or amplification rates, can be defined for the forward and backward Lyapunov vectors as discussed in
\textit{e.g.} \cite{Vannitsem1997}. In the following the stretching factor, $\chi_i (\vec x(t))$, and the local amplification
rates along the forward and backward Lyapunov vectors will be computed at the time step level of the model integration.

{\sv Since the investigation presented below is performed in an Euclidean phase space, the usual scalar (or dot) product 
is used for all the computation performed on the lengths and angles of the vectors in phase space. It is worth emphasizing that the stretching
factors along the CLVs, the CLVs themselves and the LEs are intrinsic asymptotic properties of the flow and do not depend on the 
specific norm and scalar product used. The angles on the other hand are dependent on the specific scalar product used.}

\subsection{The Construction of a Basis for the Tangent Space}

{\sv The computation of the forward and backward Lyapunov vectors is done using the classical Gram-Schmidt orthogonalisation
method. The algorithm is first based on the computation of  the amplification of a set of perturbations in
the tangent space defined at each point along the trajectory of the solution forward in time. 
This set is regularly orthonormalized using
the Gram-Schmidt approach in order to avoid numerical degeneracies between these different perturbations 
\cite{ParkerChua1989}. Here the orthonormalization is performed every integration time step (0.05 time units), but
the results are not sensitive to thie choice of the interval, provided it is kept small. Once the full trajectory and the backward
Lyapunov vectors are stored, the backward integration in time is started using the adjoint model in order to get the forward
Lyapunov vectors. 

The CLVs can then be computed as part of this backward integration at each
time step for the whole control trajectory, see also \cite{Kuptsov2012}. }   
In order to compute the basis $\{ \vec g_i\}$ one must find the intersection of two subspaces in a N-dimensional
phase space. In other words, one must find the linear combinations between the basis vectors spanning these
subspaces,

\begin{equation}
\sum_{j=1}^{N+1} r(j) \vec s(j) = 0
\end{equation}
where
\begin{eqnarray}
\vec s (j) = \vec l^-_j \,\,\, \rm{for} \,\,\, j=1,...,i \\
\vec s (j) = \vec l^+_{j-1} \,\,\, \rm{for} \,\,\, j=i+1,...,N+1
\label{lincomb}
\end{eqnarray}
or in matrix form,
\begin{equation}
\left (
\begin{array}{cccc}
s_1(1) & s_1(2) & \ldots & s_1(N+1) \\
s_2(1) & s_2(2) & \ldots & s_2(N+1) \\
\vdots & \vdots & \vdots & \vdots \\
s_N(1) & s_N(2) & \ldots & s_N(N+1) 
\end{array}
\right ) 
\left (
\begin{array}{c}
r(1) \\
r(2) \\
\vdots \\
r(N+1)
\end{array}
\right ) = 0 
\end{equation}
To solve this system with a smaller number of equations than unknowns, the singular value
decomposition {\sva is used} \cite{Trefethen1997,Press2007}.

\section{The Low-Order Atmosphere-Ocean Coupled Model}
\label{ssec:full}

The instability properties described in Sec. \ref{sec:formulation} will be studied in the context of a low-order coupled ocean-atmosphere model
displaying well-separated multiple characteristic time scales, as discussed in \cite{Vannitsem2015}. The model is described in
this section. First the original partial differential equations are briefly described, and the reduction to a low-order model is then 
outlined.

\subsection{The Equations of Motion}
\label{ssec:atmos}

The atmospheric model is based on the vorticity equations of a two-layer, quasi-geostrophic
flow defined on a $\beta$-plane \cite{Pedlosky1987}. The evolution equations in pressure coordinates are
\begin{eqnarray}
\frac{\partial}{\partial t} \left( \nabla^2 \psi^1_a \right) + J(\psi^1_a, \nabla^2 \psi^1_a) + \beta \frac{\partial \psi^1_a}{\partial x}
& = & -k'_d \nabla^2 (\psi^1_a-\psi^3_a) + \frac{f_0}{\Delta p} \omega, \nonumber \\
\frac{\partial}{\partial t} \left( \nabla^2 \psi^3_a \right) + J(\psi^3_a, \nabla^2 \psi^3_a) + \beta \frac{\partial \psi^3_a}{\partial x}
& = & +k'_d \nabla^2 (\psi^1_a-\psi^3_a) - \frac{f_0}{\Delta p}  \omega \nonumber \\  
& & - k_d \nabla^2 (\psi^3_a-\psi_o); 
\label{eq:atmos}
\end{eqnarray}
here $\psi^1_a$ and $\psi^3_a$
are the streamfunction fields at $p_1=250$ and $p_3=750$ hPa, respectively, and $\omega = dp/dt$ is the vertical velocity.
{\sva $f_0$ is the Coriolis parameter and $\beta = df/dy$ its meridional gradient, at latitude $\phi_0= 45^\circ$ N.}  
The coefficients $k_d$ and $k'_d$
multiply the surface friction term and the internal friction between the layers, respectively, while $\Delta p = 500$ hPa is the pressure
difference between the two atmospheric layers. An additional term has been introduced in this system in order to account for the presence of a
surface boundary velocity of the oceanic flow defined by $\psi_o$. 

The ocean component is based on the reduced-gravity, quasi-geostrophic shallow-water model on a $\beta$-plane{\sva, describing the dynamics of a fluid layer of constant
density superimposed on a quiescent deep layer \cite{Dijkstra2005}}:
\begin{equation}\label{eq:QGSW}
\frac{\partial}{\partial t} \left( \nabla^2 \psi_o - \frac{\psi_o}{L_R^2} \right) + J(\psi_o, \nabla^2 \psi_o) + \beta \frac{\partial \psi_o}{\partial x}
= -r \nabla^2 \psi_o + \frac{{\mathrm{curl}}_z \tau}{\rho h}.
\end{equation}
where $\psi_o$ is the
streamfunction in the model ocean's upper, active layer,
$\rho$ the density of water of the upper layer, $h$ the depth of
this layer, $L_R$ the reduced Rossby deformation radius, $r$ a friction coefficient at the bottom of the
active layer, and ${\mathrm{curl}}_z \tau$ is the vertical component of the curl of the wind stress.
It is assumed that the wind stress is given by $(\tau_x, \tau_y)=C (u-U,v-V)$ --- where  {\svc $(u = -\partial \psi^{1,3}_a/\partial y, v = \partial \psi^{1,3}_a/\partial x)$ } are the 
horizontal components of the geostrophic wind{\svc, and $U= -\partial \psi_o/\partial y$ and $V= \partial \psi_o/\partial x$ are the horizontal velocity fields within the ocean}. 
A drag coefficient defined as $d = C/(\rho h)$ characterizes the strength of the mechanical coupling between the ocean and the atmosphere.

\subsection{Thermodynamic Equations}
\label{ssec:ocean_temps}

The ocean temperature, {\sva, $T_o$,} is advected as a passive scalar by the ocean currents and is strongly coupled to the atmospheric temperature through radiative and heat
exchanges:
\begin{equation}\label{eq:heat_oc}
\gamma_o ( \frac{\partial T_o}{\partial t} + J(\psi_o, T_o)) = -\lambda (T_o-T_a) + E_R.
\end{equation}
with
\begin{equation}\label{eq:fluxes_oc}
E_R = -\sigma_B T_o^4 + \epsilon_a \sigma_B T_a^4 + R_o.
\end{equation}
In Eqs.~(\ref{eq:heat_oc}) and (\ref{eq:fluxes_oc}) above, $E_R$ is the net radiative flux at the ocean surface, $R_o$ is
the shortwave radiation entering the ocean, $\epsilon_a$ the emissivity of the atmosphere, $\sigma_B$ the Stefan-Boltzman constant, $\gamma_o$
the heat capacity of the ocean, and $\lambda$ is the
heat transfer coefficient between the ocean and the atmosphere
that combines both the latent and sensible heat fluxes. It is assumed that the combined heat transfer is
proportional to the temperature difference between the atmosphere and the ocean.

Following the quasi-geostrophic approximation, and using the equation of state of ideal gases $p=\rho_aRT_a$, where $R$ is the gas constant, the temperature of the atmosphere can be expressed as $T_a = - p f_0/R \, \partial \psi_a/\partial p$. Since our model includes only two vertical layers, the temperature is defined only at the intermediate layer $(p_1+p_3)/2$ as $T_a = - p f_0/R \, (\psi_a^3-\psi_a^1)/(\Delta p)$. 
The equation for {\sva $T_a$} is given by
\begin{equation}\label{eq:heat_atm}
\gamma_a ( \frac{\partial T_a}{\partial t} + J(\psi_a, T_a) -\sigma \omega \frac{p}{R}) = -\lambda (T_a-T_o) + E_{a,R}
\end{equation}
with
\begin{equation}\label{eq:fluxes_atm}
E_{a,R} = \epsilon_a \sigma_B T_o^4 - 2 \epsilon_a \sigma_B T_a^4 + R_a.
\end{equation}

In Eqs. \ref{eq:heat_atm}-\ref{eq:fluxes_atm}, $\sigma$
is the static stability, taken to be constant. 
It is straightforward to combine Eqs.~(\ref{eq:heat_atm}) and (\ref{eq:atmos}),
as done when deducing the quasi-geostrophic potential vorticity equation \cite{Vallis2006}. {\sv The atmospheric temperature is
not per se a prognostic equation in the model and is deduced using the diagnostic relation mentioned above, but the
impact of the radiative forcing is felt in the vorticity field via the vertical velocity, $\omega$, appearing in both Eqs.~(\ref{eq:heat_atm}) and (\ref{eq:atmos}).}

\subsection{ Low-order model formulation}
\label{sec:low-dim}

In order to build a low-order model version, the fields are expanded in Fourier series
 and truncated at a
minimal number of modes that still captures key features of observed behavior. Both linear and nonlinear terms in the equations of motion are then
projected onto the
phase subspace spanned by the modes retained, by using an appropriate scalar product.
For the closed ocean basin one uses only sine functions in order to enforce no-flux conditions at the boundaries.
For the atmosphere, no-flux boundaries are assumed in the meridional direction,while periodic boundary conditions are taken in the longitudinal direction
as discussed in \cite{Reinhold1982,Vannitsem2013,Vannitsem2014}. {\sv The radiative input is given by 
\begin{subequations}\label{eq:rad}
\begin{align}
& R_{\rm a} = R_{{\rm a},0} + \delta R_{\rm a}, \label{rad_atm} \\
& R_{\rm o} = R_{{\rm o},0} + \delta R_{\rm o}, \label{rad_oc}
\end{align}
\end{subequations}
with $R_{{\rm o},0}$ and  $R_{{\rm a},0}$ the averaged shortwave radiative forcings over the domain considered. }
For the latitudinal variation of radiative fluxes, we will assume
that it is proportional to a cosine function of latitude,
\begin{equation}
\delta R_o(t)=4 \delta R_a(t) = C_o \sqrt{2} cos(\pi y / L_y)
\end{equation}
{\sv where {\sva $L_y$ is the meridional dimension of the rectangular horizontal domain over which the equations are integrated}, and we assume that the short-wave radiative input 
within the atmosphere is 1/4 the one entering the ocean. {\sva The aspect ratio, $n=2 L_y /L_x$, of the domain is set to $n=1.5$.} } 
We retain 8 modes for the ocean and 10 modes for the atmosphere \cite{Vannitsem2015}.

In order to overcome the problem of the quartic terms in the radiative fluxes,
we will take advantage of the small amplitude of temperature anomalies, as compared with a reference temperature, in order to linearize
these terms. 

The equations are nondimensionalized by scaling horizontal distances by $L, (x'=x/L, y'=y/L)$, time $t$ by
$f_0^{-1}$, the vertical velocity $\omega$ by $f_0 \Delta p$ and  the atmospheric and oceanic streamfunctions $\psi_a$ and $\psi_o$ by $L^2 f_0$.
The parameters are
also rescaled as
$$2 k = k_d/f_0, k' = k'_d/f_0$$  $$\beta' = \beta L/f_0$$
$$\gamma= - L^2/L_R^2$$ $$r'=r/f_0$$ $$\delta = d/f_0 = C/(\rho h f_0)$$

{\sv The typical space and {\sva inverse-}time scales of our model are $L_y=\pi L = 5000$ km and $f_0=0.0001032$ s$^{-1}$. The parameter values are 
{\sva set in the present work to }
$2k=k'=0.04$, $r'=0.000969$, $\beta'=0.2498$, $\gamma=-1741$, {\sva $C_o=350$W m$^{-2}$} and $d=10^{-8}$ s$^{-1}$ (except in some experiments of Section 5 for which $d$ has been
modified). }
The temperatures are also made nondimensional, $T'_o = T_o R/(f_o^2 L^2) $ and $T'_a = T_a R/(f_o^2 L^2) $
{\sv Note that the friction parameter within the ocean is 
{\svc difficult to assess from real ocean data,} as discussed in \cite{Nese1993},
but one can argue that it should be proportional to the amplitude of the velocity, $C_D |V|$,  which is two to three orders of magnitude smaller within the
ocean than within the atmosphere. This would suggest -- provided we assume that the friction coefficient, $C_D$, is similar in both sub-systems -- 
that friction within the ocean would be much less important than within the atmosphere, due to the natural inertia of the ocean.} 

{\sva The dynamics of the low-order model is thus described by a set of 36 ordinary differential equations with quadratic nonlinearities.}
 
\section{Lyapunov Properties of the System}
\label{sec:results}

\subsection{Geometry of the Tangent Space}
\label{subsec:Lyapunov}

Information in phase space on the uncertainties of the future evolution of the flow is an essential ingredient when forecasts are issued.
This has long be recognized in particular in meteorological applications and is at the origin of the development of probabilistic forecasts. 
A central ingredient in the development of this approach is to compute the local instability of the flow either in physical or phase space
(\cite{Kalnay2003}). As discussed in Section (\ref{sec:formulation}), the finite time Lyapunov exponents (FTLEs) can be evaluated based on either the forward, backward and covariant Lyapunov vectors. We define as $\sigma_{j,X}^0(k)$  the estimate of the $j^{th}$ instantaneous FTLE at time step $k$ computed according to the geometrical construction $X$, where $X=C$, $F$, $B$ indicates covariant, forward, and backward Lyapunov vectors, respectively. Therefore, we refer to $\sigma_{j,C}^0(k)$, $\sigma_{j,F}^0(k)$, and $\sigma_{j,B}^0(k)$ as a FTCLE, FTFLE, and FTBLE, respectively.  

The (infinite time) LEs are obtained by averaging the FTLEs over an infinitely long trajectory, so that from the FTCLEs, FTFLEs, and FTBLEs we derive by averaging the CLEs, FLEs, and BLEs, respectively. As mentioned in Sec. \ref{sec:formulation}, the theory suggests that $$\lim_{M \rightarrow \infty} \frac{\sum_{k=1}^M \sigma_{j,X}^0(k)}{M}=\lim_{M\rightarrow \infty} \sigma_{j,X}^M(k)= {\sigma}_j, \quad\forall j,\forall X=C,F,B  $$
where  ${\sigma}_j$ is the $j^{th}$ LE and $\sigma_{j,X}^M(k)$ is the $j^{th}$ FTLE obtained by averaging over $M$ time steps starting at time step $k$. Note that the theory does not impose that the moments larger than one of the probability distribution of the instantaneous (and non instantaneous) FTCLEs, FTFLEs, and FTBLEs should be the same{\sv. {\sva This will be analyzed in detail in the next section.} Instead, the asymptotic Lyapunov exponents computed using the different methods should be identical within numerical precision}{\sva, as illustrated below.}  

\begin{figure}
\centering
a){\includegraphics[width=70mm]{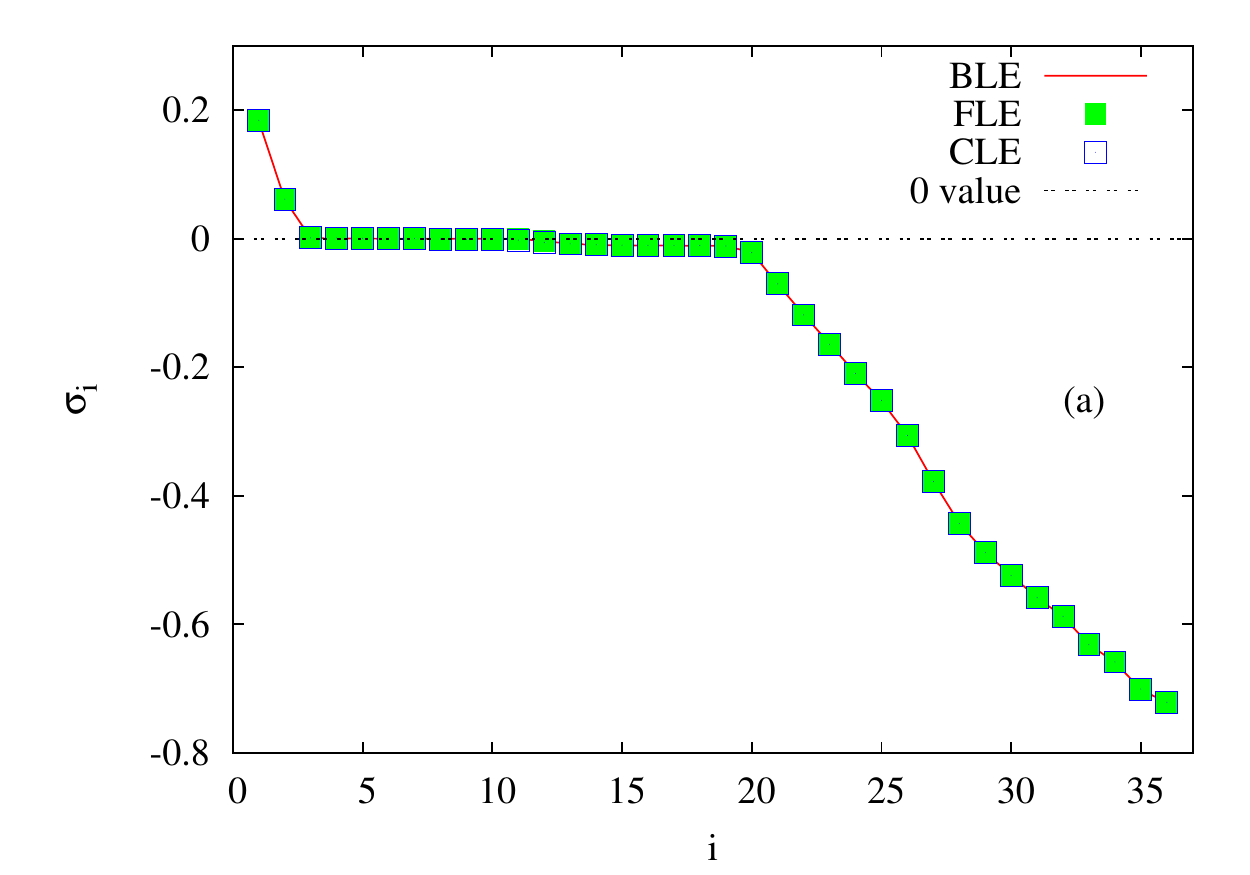}}
b){\includegraphics[width=70mm]{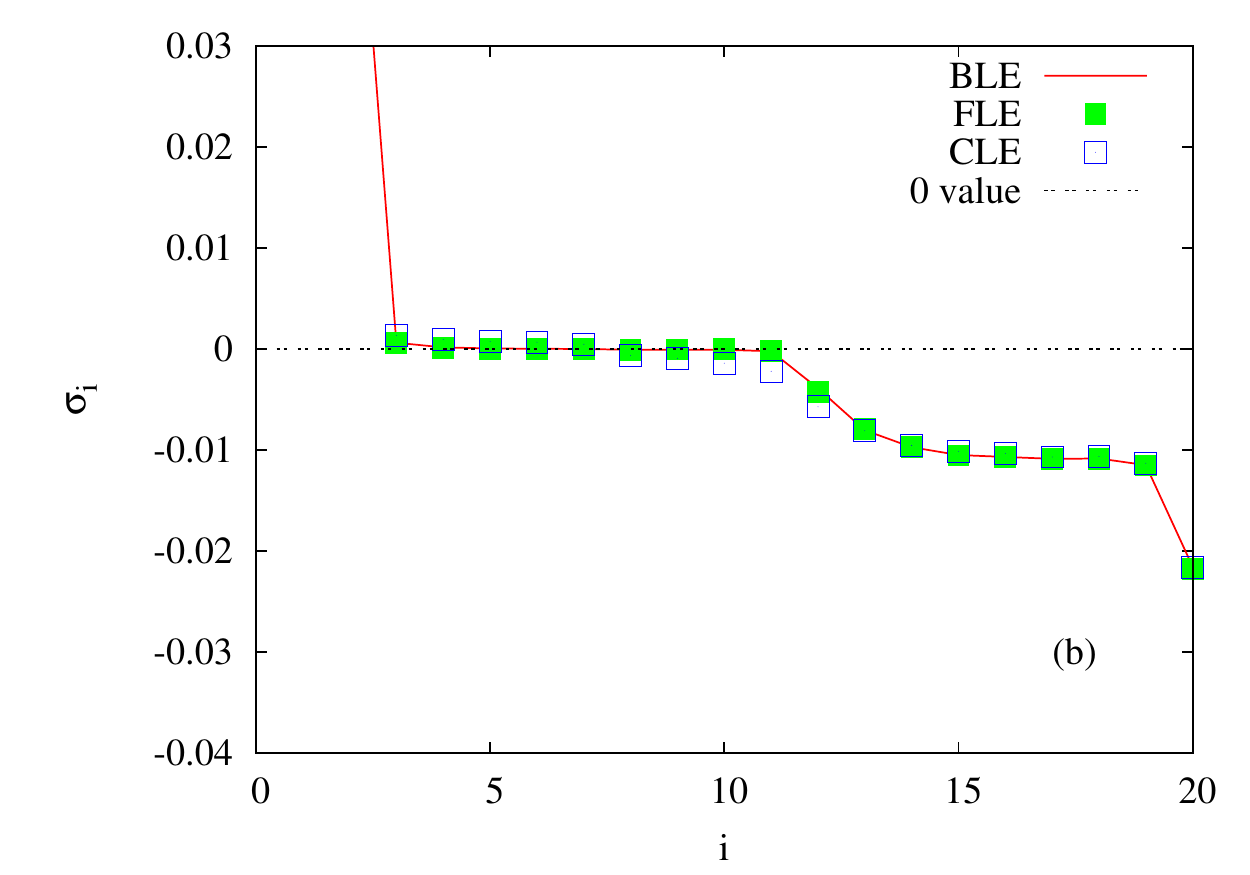}}
\caption{Lyapunov {\sva exponents, given in day$^{-1}$,} based on the forward integration (FLEs, red line), backward integration (BLEs, full green squares), and on the CLVs (CLEs, empty blue squares). Panel (a): the full spectrum of LEs. Panel (b) a zoom around 0. Parameters' value: $C_o=350$ W m$^{-2}$ and $d=1 \times 10^{-8}s^{-1}$.}
\label{Lyap_d1x10-8}
\end{figure}

Figure \ref{Lyap_d1x10-8}a displays all the LEs (also referred to as Lyapunov spectrum) as obtained with a long time integration of the coupled ocean-atmosphere model of Sec. \ref{ssec:full} for the two control parameters, $C_o=350$ W m$^{-2}$ and $d=10^{-8}$ s$^{-1}$. In the simulation, averages are performed over the whole considered integrations, which cover 90,000, 56,000 and 56,000 days for the backward, forward and covariant Lyapunov vectors, respectively, thus giving the estimates for the BLEs, FLEs, and CLEs, respectively. The model in this configuration is chaotic with apparently two positive LEs (group {\it a}),  a set of 18 LEs very close to 0 (group {\it c}), and, subsequently, 16 negative LEs (group {\it b}). The differences  between the estimates obtained using the three geometric constructions come from the fact that a portion of the backward model integration is discarded in order to get convergence of the forward Lyapunov vectors. The three spectra are very
close to each other, thus indicating that the algorithm is correctly implemented and that the lenght of the integration is sufficient, although some unavoidable uncertainty remains for the exponents close to 0, which are related to ultralong time scales, see Fig. \ref{Lyap_d1x10-8}b.     

{\sv These Lyapunov spectra are obtained with a specific set of parameters. Changing parameter values could lead to strong modifications of
the dynamical properties. Two parameters playing an important role are $C_o$ and $d$ which could 
considerably affect the Lyapunov spectrum, as discussed in details in \cite{Vannitsem2015}. Beside the
geometric parameters ($n$, $L$, $f_0$, $\beta$, $\gamma$), which are usually kept fixed, three other parameters could lead
to important modifications in the dynamics, namely $k_d$, $k_d'$ and $r$. These parameters control the linear dissipation within the model, and therefore
the sum of all LEs. If $k_d$ and $k_d'$,which determine the dissipation within the atmosphere are modified, say by one order of magnitude, 
groups {\it a} and {\it b} of the spectra are considerably modified, with a larger stability for $k_d$ and $k_d'$ large. If $r$, which controls the dissipation within the 
ocean, is modified by one or two orders of magnitude, all groups are essentially unaffected mostly because dissipation within the ocean is much smaller than
within the atmosphere. Group {\it c} is quite insensitive to dissipation parameters. 
} 




\subsubsection{Variance of the CLVs}
\begin{figure}[ht]
\centering
{\includegraphics[trim=1cm 6cm 1cm 6cm,clip,scale=0.6]{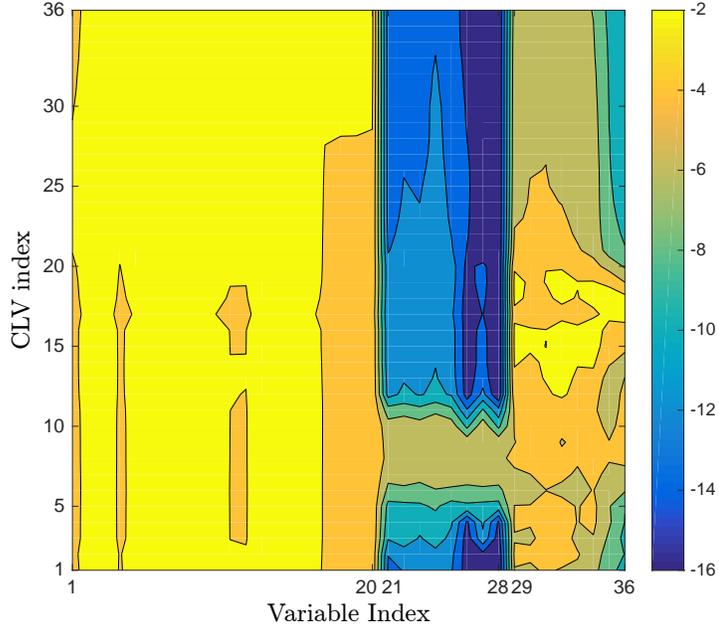}}
\caption{Values of the time-averaged and normalized variance of the CLVs as a function of the variables of the model ($\log_{10}$ scale).  The 20 first modes corresponds to the variables of the atmosphere, the next 8 ones
to the dynamics within the ocean and the last 8 ones to the temperature within the ocean. Parameters' value: $C_o=350$ W m$^{-2}$ and $d=1 \times 10^{-8}s^{-1}$. {\sv The Euclidean norm is used for all CLVs, and their squared norm is normalized to 1.}}
\label{Lyap_d1x10-8b}
\end{figure}

We investigate the physical meaning of the instabilities described by the LEs by first looking at  the averaged variance of  the corresponding CLVs on the variables of the model (Fig. \ref{Lyap_d1x10-8b}). The Euclidean norm is used for all CLVs,
and their squared norm is normalized to 1. The 10 first variables correspond to the barotropic atmospheric streamfunction, the next 10 variables to the baroclinic atmospheric 
streamfunction, the following 8 variables to the ocean velocity field, and the 8 final ones to the ocean temperature field.

As expected, the variance of the CLVs is dominantly localized in the atmospheric variables, which provide the dominant component contributing to the dynamics of the system. In particular, the two unstable CLVs (corresponding to the first two LEs) are dominantly atmospheric, and so are the very stable CLVs 21-36, featuring large and negative LEs. 

For CLVs 1 and 2, the oceanic thermodynamic variables have also a substantial projection, thus implying that instabilities result from coupled atmospheric-oceanic modes, where the coupling, though, does not involve the oceanic dynamic variables. 

For  CLVs 21-36, the projection on the oceanic variables is extremely small, and becomes smaller and smaller as we explore higher and higher values of the LEs. Also in this case, the ocean contributes almost exclusively through its faster thermodynamic variables, while the projection on the dynamical variables is almost absent.

When considering CLVs with LEs close to zero (3-20), we have much larger projections on the oceanic variables. For CLVs 12-20, the variance projected on the oceanic thermodynamic variables is comparable to the projection on the atmospheric one, thus indicating a very strong level of coupling between the two geophysical fluids. Instead, the role of the dynamic oceanic variables is entirely negligible.

CLVs 3-11 provide a very interesting piece of information. In this case - and only for these CLVs - also the oceanic dynamic variables play an important role. These are the only group of CLVs where the variance is distributed across all variables, and correspond to the modes featuring the longest time scales; note that  LEs 3-11 are in absolute value much smaller than LEs 12-20, see Fig. \ref{Lyap_d1x10-8}b.


\subsubsection{Statistics of the Angles between the CLVs and Time Scale Separation}

By construction, the CLVs span the tangent space of the system.  
Nonetheless, it is important to check the statistics of the angles between them in order to assess the degree of separation of the subspaces corresponding to different modes (or combination of) and, conversely, whether degeneracies occur in the form of near-tangencies \cite{Yang2009,Takeuchi2011}. 
The angles have been computed with reference to the Euclidean scalar product.

\begin{figure}
\centering



a){\includegraphics[trim=0 0 0 1cm,clip,width=60mm]{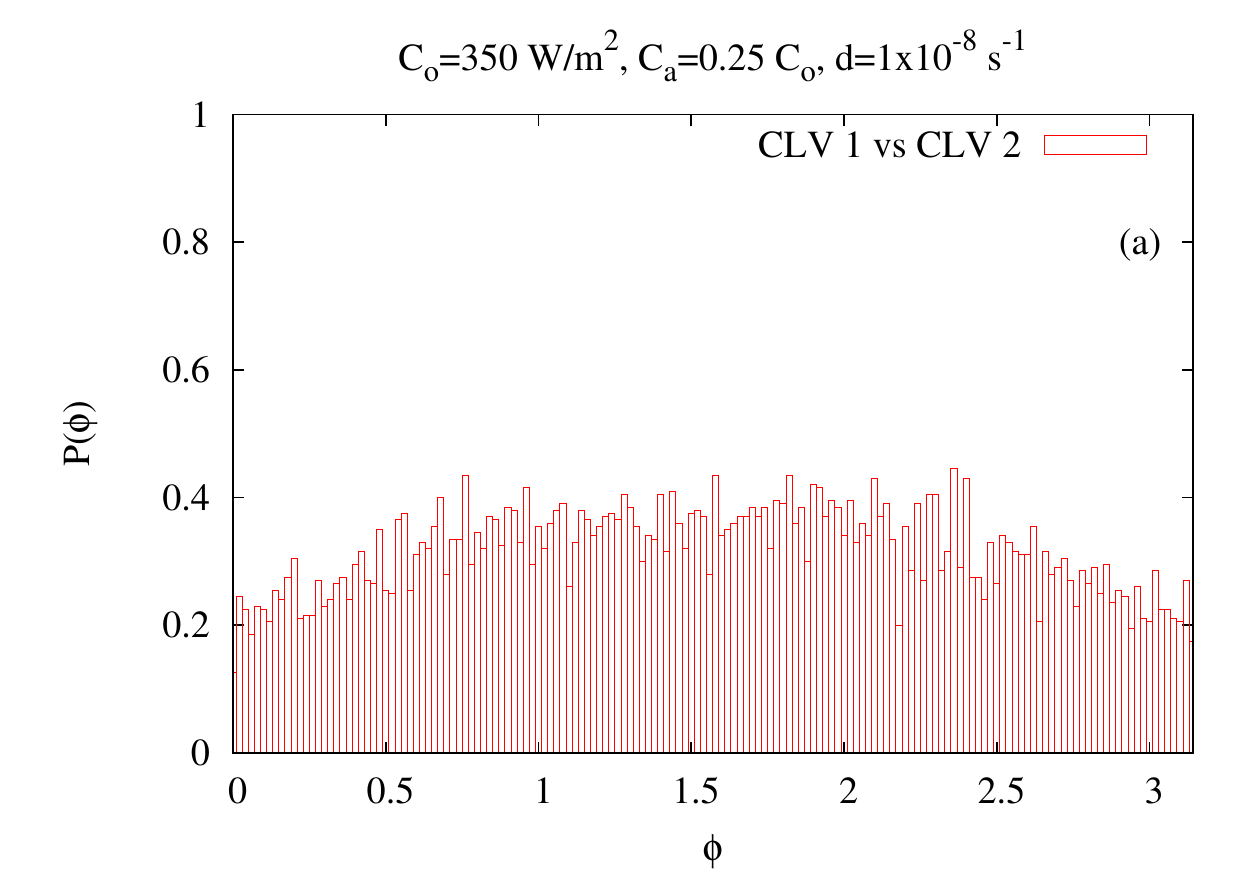}}
b){\includegraphics[trim=0 0 0 1cm,clip,width=60mm]{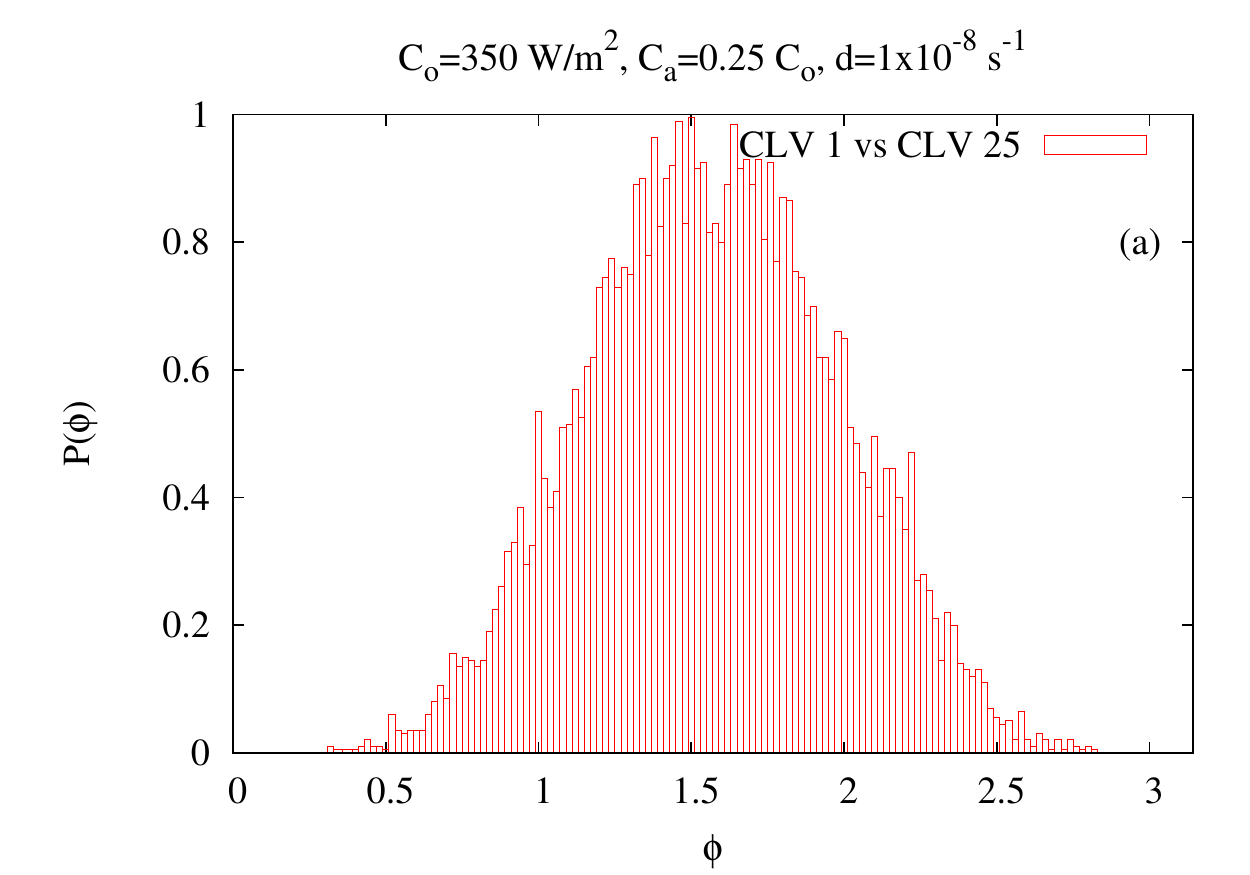}}

c){\includegraphics[width=60mm]{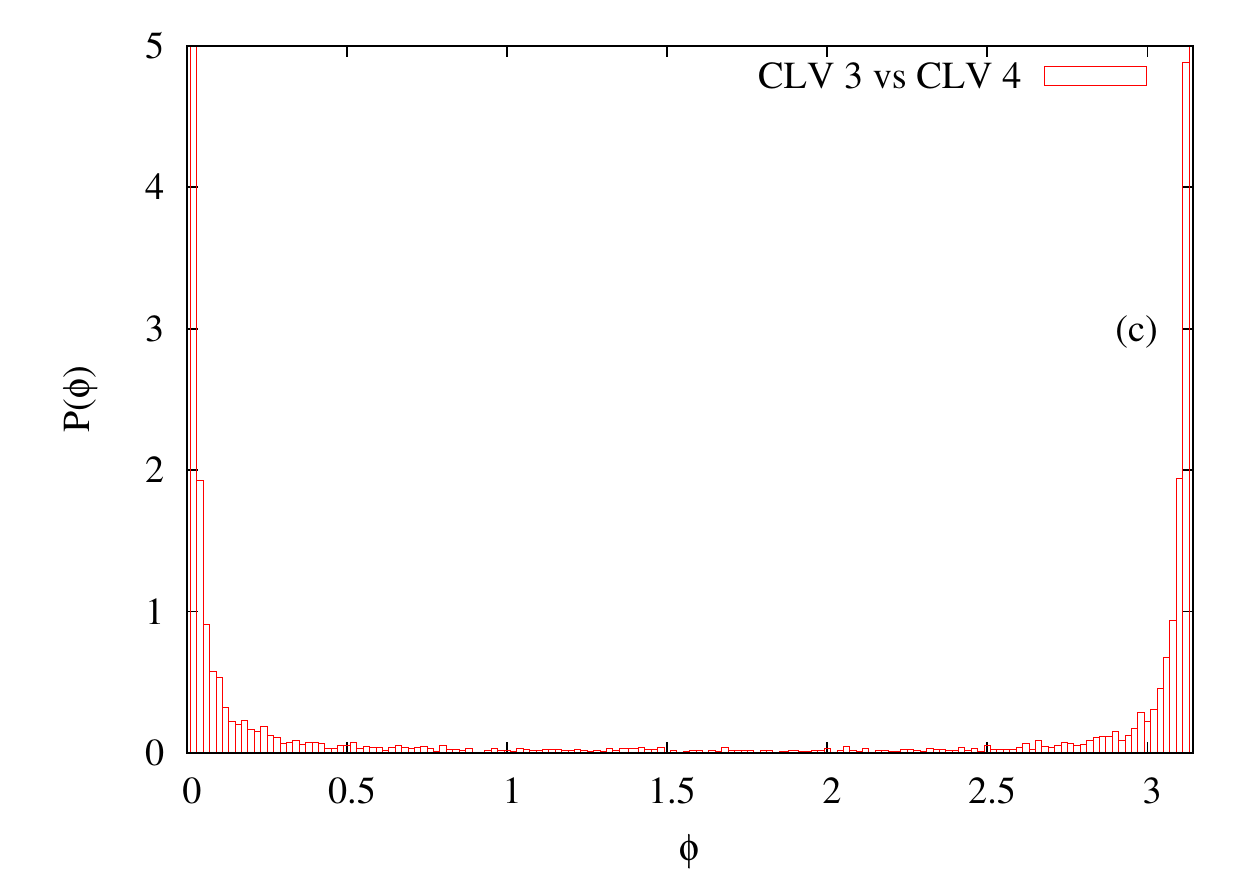}}
d){\includegraphics[trim=0 0 0 1cm,clip,width=60mm]{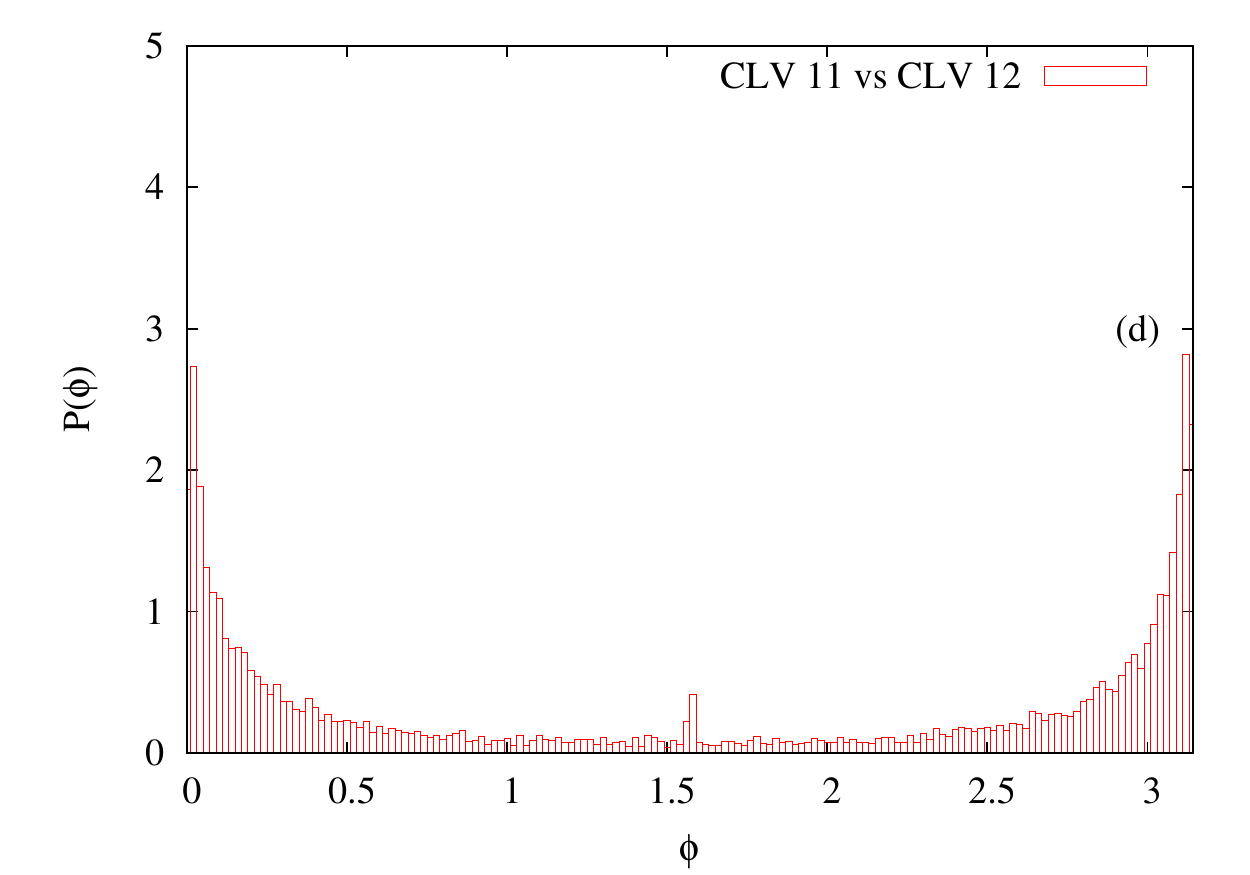}}

e){\includegraphics[width=60mm]{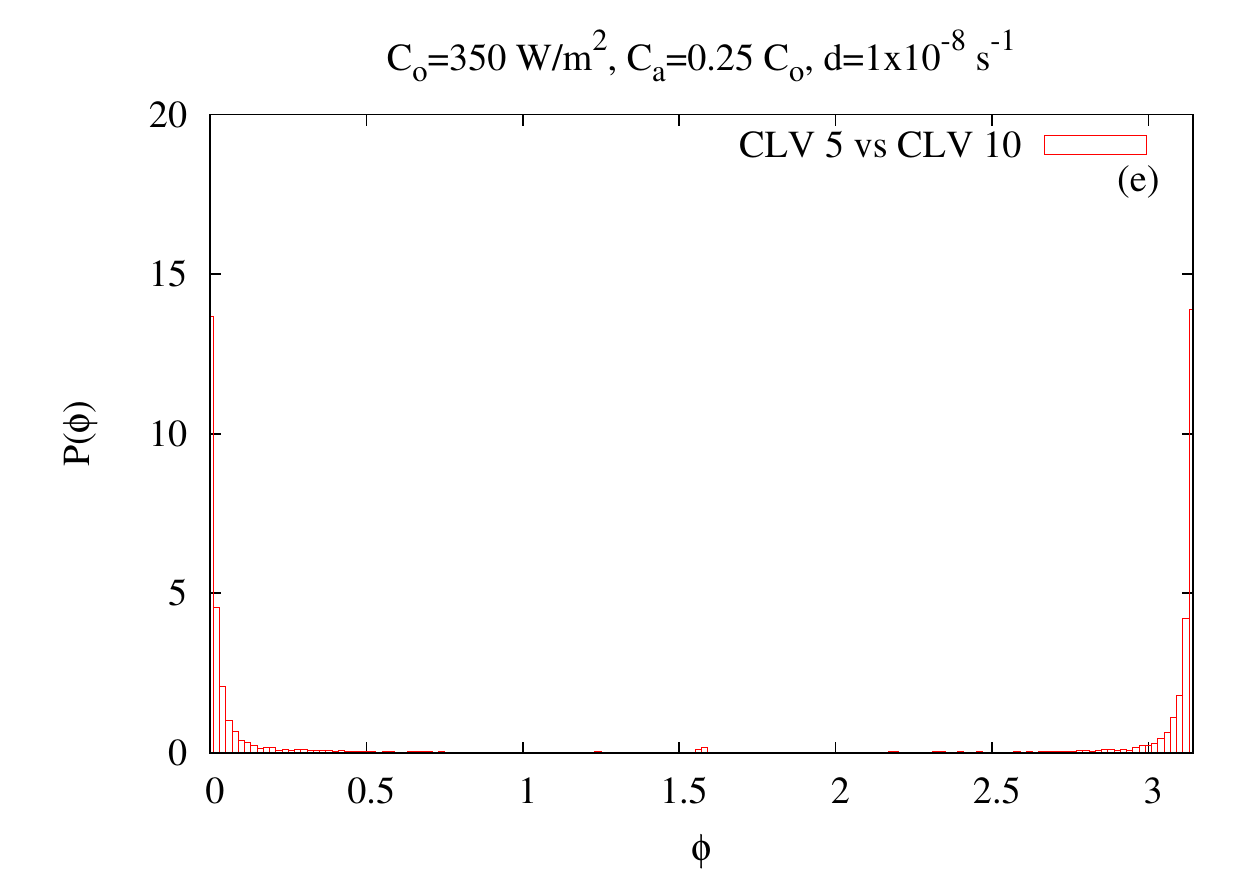}}
f){\includegraphics[width=60mm]{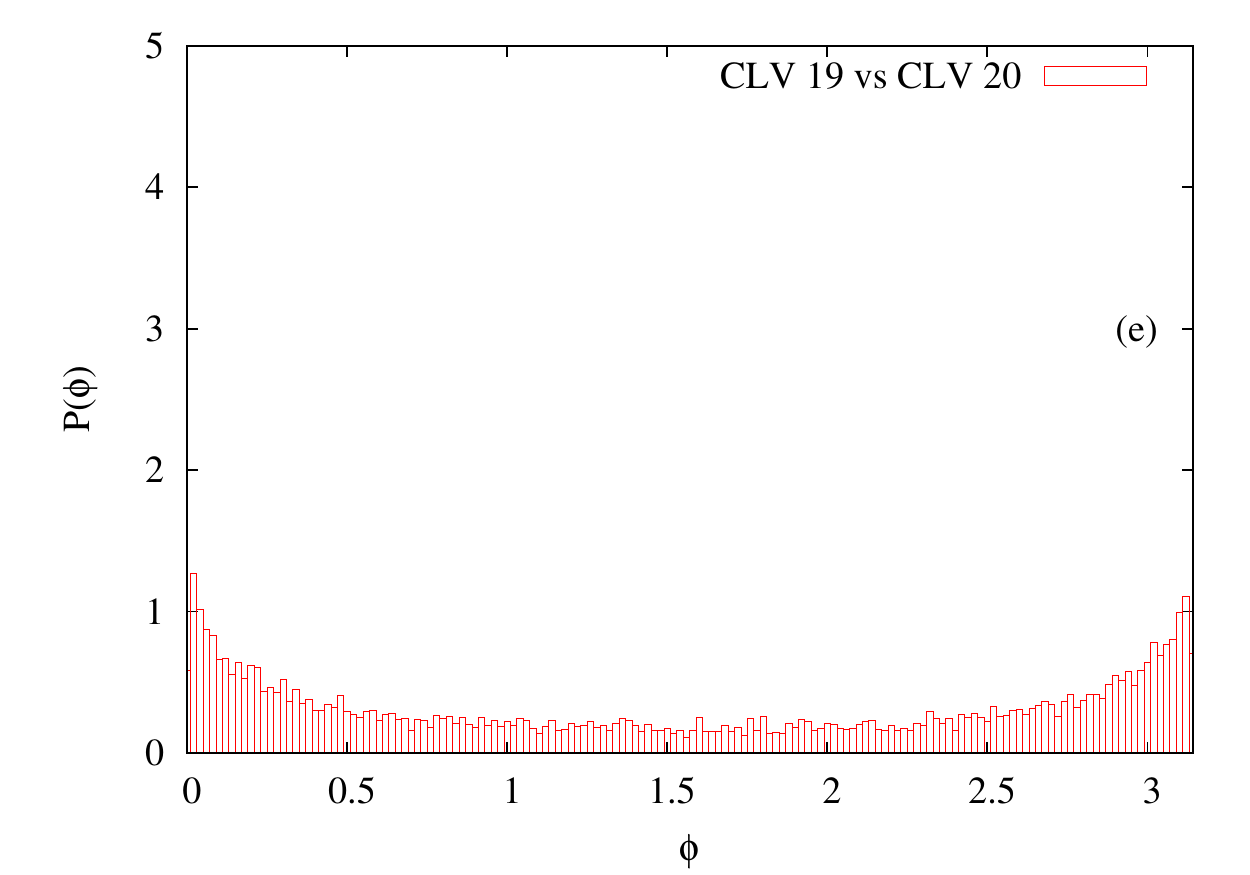}}

g){\includegraphics[width=60mm]{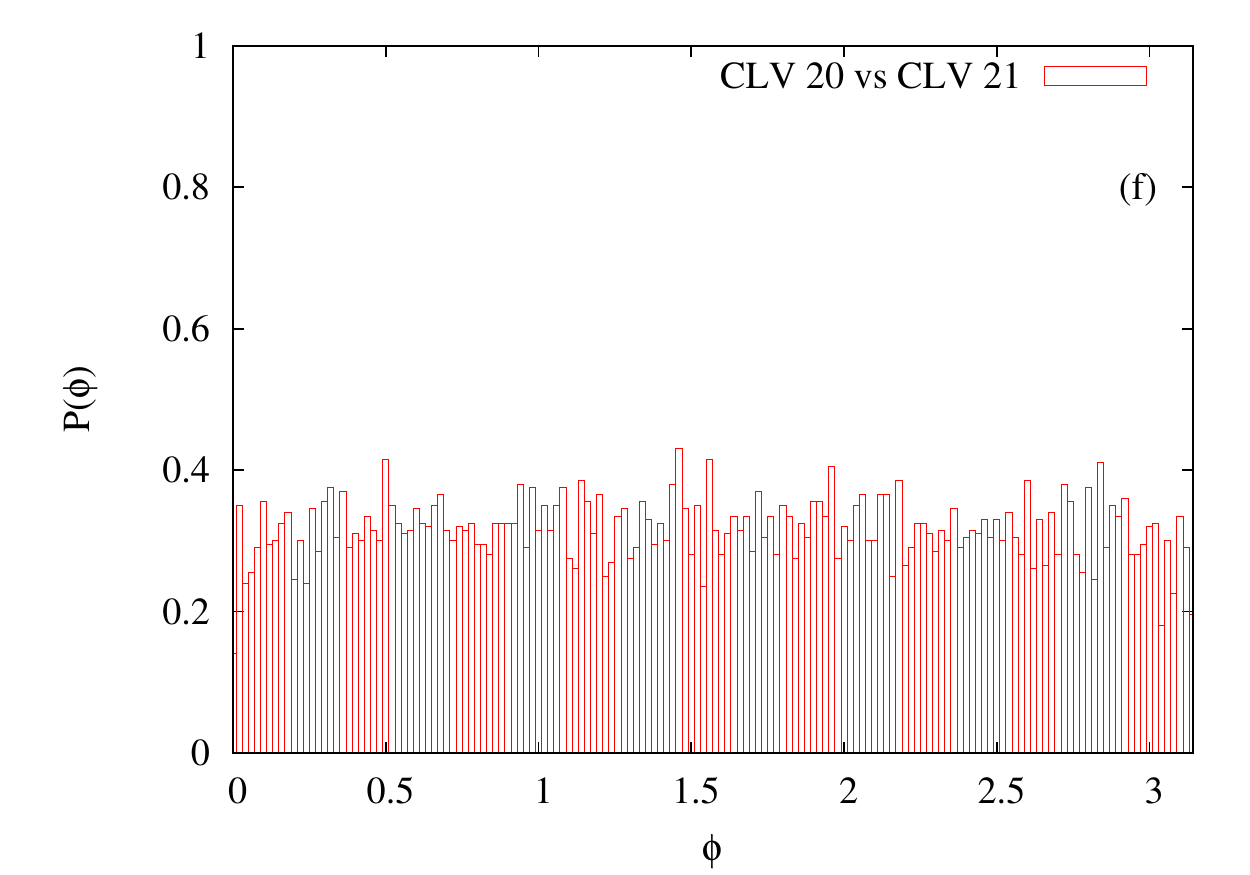}}
h){\includegraphics[width=60mm]{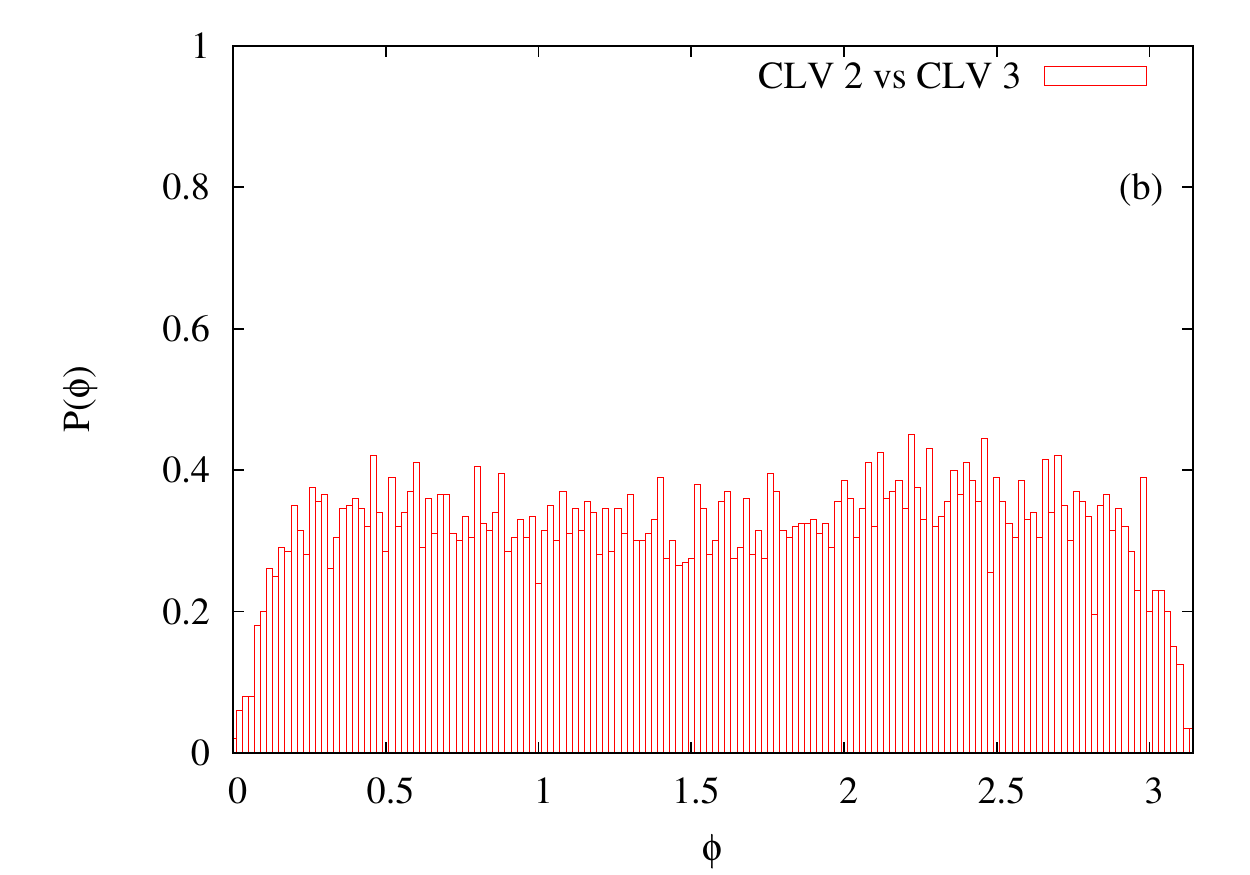}}
\caption{Nonparametric estimates of the probability densities of angles $\phi$ between pair of CLVs, (a) CLV 1 vs CLV 2, (b) CLV 1 vs CLV 25, (c) CLV 3 vs CLV 4, (d) CLV 5 vs CLV 10, (e) CLV 11 vs CLV 12, (f) CLV 19 vs CLV 20, g) CLV 20 vs CLV 21,
(h) CLV 2 vs CLV 3. Note that the angle runs from $0$ to $\pi$ because statistics of $\phi$ and $2\pi-\phi$ are combined. Parameters' value: $C_o=350$ W m$^{-2}$ and $d=1 \times 10^{-8}s^{-1}$.}
\label{angles2}
\end{figure}

Figures \ref{angles2}a-\ref{angles2}h display the probability density of the angles for different combinations of CLVs, chosen among the qualitatively homogeneous groups {\it a}, {\it b}, and {\it c} described above.

Panel (a) shows the probability distribution of the angles between CLV 1 and CLV 2, which covers the entire range between 0 and $\pi$. Similar results are found when evaluating the probability distribution of the angles between two CLVs belonging to group {\it b}  (not shown). Broadly speaking, the larger the index difference between the considered CLVs, the more the angles are localized around $\pi/2$, reflecting the progressive \textit{independence} between the vectors due to increasing time scale separation \cite{Yang2009, Takeuchi2011}; see \textit{e.g.} the distribution of angles between CLV 1 and CLV 25 (panel b). This is especially pronounced when considering CLVs with large index, which suggests that these modes play a minor role in the dynamics of the system, as a result of their strong (atmospheric) dissipative behaviour, already discussed in the previous subsection.   
 
Instead, for all CLVs belonging to group {\it c} we find an extreme degree of geometrical degeneracy, in the sense that their angles are almost invariably very close to $0$ or $\pi$, see panels (c-f) for illustrative examples. The coupling is even stronger inside the two subgroups described above, projecting more on  dynamical or thermodynamical oceanic variables, respectively.  As a result, as soon as weakly nonlinear effects are considered, these modes are coupled, in the sense that small perturbation in one CLV propagates to all the others. This is not the case when the density of angles is finite (or a fortiori when it vanishes) near $0$ or $\pi$ \cite{Yang2009,Takeuchi2011}. 
The existence of exact tangency between two vectors is clearly norm-independent, and the presence of very frequent occurrence of near tangency is only mildly norm-dependent.

Note that, when we consider the statistics of angles between CLV 20 (group {\it c}) and CLV 21 (group {\it b}), we recover the typical broad distribution of angles, where no degeneracy is found (see panel g), even if the corresponding LEs are pretty similar. It is useful to note the  changeover between panel (f) and panel (g), where consecutive CLVs are studied, which clearly indicates that there is qualitative change in the properties of the vectors for index values larger than 20.  Similar observation can be made when comparing panel c) (statistics of the angles between CLVs 3 and 4) with panel h) (statistics of the angles between CLVs 2 and 3). Also in this case, we have a clear signature of the fundamental change in the properties of the CLVs when passing from index 2 to index 3.  LEs 2, 3, 4 and LEs 19, 20, 21 are all extremely close to each other: our finding confirms that similarity between the values of the LEs is not a sufficient condition to assess dynamical degeneracy between two corresponding CLVs, because very diverse dynamical processes can lead to similar amplifications rates.


Note that by changing one parameter of the model it is possible to see even more clearly the link between the presence of ultralong time scales and the difficulties in estimating of the properties of the tangent space. As discussed in details in \cite{Vannitsem2015}, strongly different dynamics can emerge
when $d$ is increased. For $d=6 \times 10^{-8} s^{-1}$, the chaotic attractor develops around a very slow periodic orbit with a period of about 20 years. and, while the system has sensitive dependence on initial conditions when very long time horizons are considered, in substantial parts of the attractor the largest FTLE vanishes. 
The convergence of the LEs is more difficult to obtain, especially for the near-zero exponents, where significant
differences between the three methods of computation are apparent (not shown). 



This reinforces the idea that in the case of a system displaying a ultralow frequency variability, the computation of the CLVs close to 0 is extremely tricky. Longer integrations are therefore needed with a very important cost in terms of disk space since the backward Lyapunov
vectors must be stored during the whole integration. The approaches proposed in \cite{Wolfe2007} and in \cite{Ginelli2007} are therefore worth considering
at the expenses of the set of exponents that can be effectively computed. This seems very relevant in terms of applications on seamless prediction for weather \textit{and} climate time scales.       


\subsubsection{Hyperbolicity, Nonuniform Hyperbolicity, and Partial Hyperbolicity}

As discussed above, the system can be divided into three main dynamical components: {\it a}, the unstable dynamics associated with two positive LEs (1-2), {\it b}, the stable dynamics associated with the sixteen negative LEs (21-36), {\it c}, the remaining eighteen near-zero and zero LEs (3-20). 

Therefore, the dynamics of the system can be seen as the product of something essentially mixing associated with the unstable (component {\it a}) and stable (component {\it b}) manifold, times a complex, highly geometrically degenerate and high-dimensional weakly chaotic or non-chaotic dynamics with slowly mixing (if at all) behaviour (component {\it c}). Note that, on the time scales of relevance for the components  {\it a} and {\it b}, it makes virtually no difference whether weak chaos with ultralong time scales, slow decay of correlations, or quasi-periodic behaviour is realised in component {\it c}. Clearly, the special properties of the component {\it c}) result  from the presence of the longer time scales associated with the ocean dynamics and thermodynamics. This can be seen as an evidence of how multiscale system are much harder to characterize dynamically and, in numerical terms, are associated with stiff problems. 

Since we find many near-zeros LEs, the mathematical framework of nonuniformly hyperbolic systems \cite{Barreira2005} - and let alone of Axiom A systems \cite{Ruelle1989}, which are hyperbolic on the attractor - seems not useful for describing the properties of our model. 

It seems instead useful to introduce the mathematical framework provided by the partially hyperbolic systems \cite{Hasselblatt2005}, which generalize hyperbolic system by accommodating the so-called center directions in the tangent space - corresponding to our component c - where nothing contracts as rapidly as in the stable directions and nothing expands as fast as in the unstable directions. 

Nonetheless, we need to remind that also such a framework does not fit perfectly to our case as the FTLEs of the stable directions and some of those of the unstable directions have sometimes value of the opposite sign with respect to the corresponding LEs (see below), thus breaking the hyperbolicity assumption requested by the partially hyperbolic system for the stable and unstable directions. We will discuss these aspects in the conclusions. 

\subsection{Fluctuations of the Finite Time Lyapunov Exponents}
\label{subsec:fluctuations}

\begin{figure}[ht]
\centering
a){\includegraphics[trim= 1cm 0cm 0cm 0cm, clip, width=70mm]{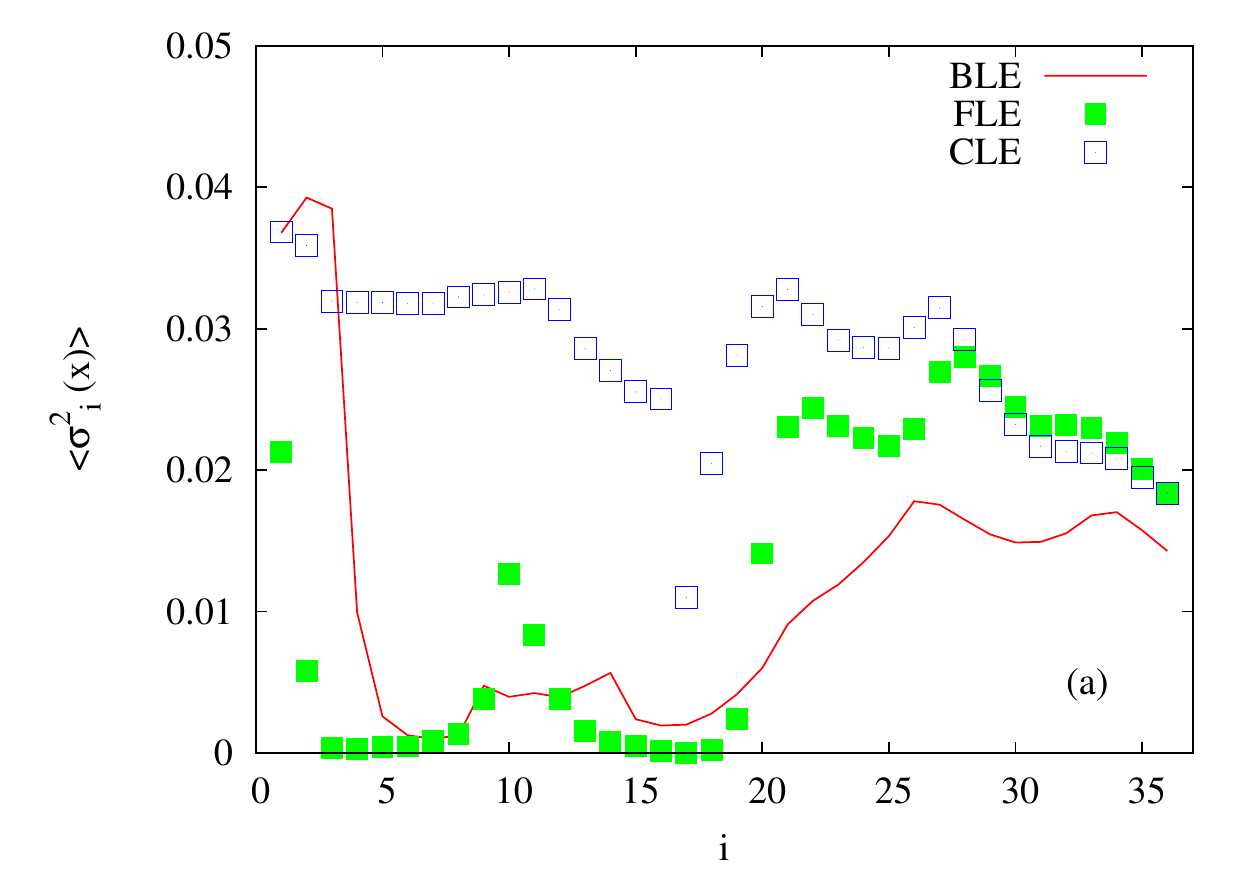}}
b){\includegraphics[trim= 1cm 0cm 0cm 0cm, clip, width=70mm]{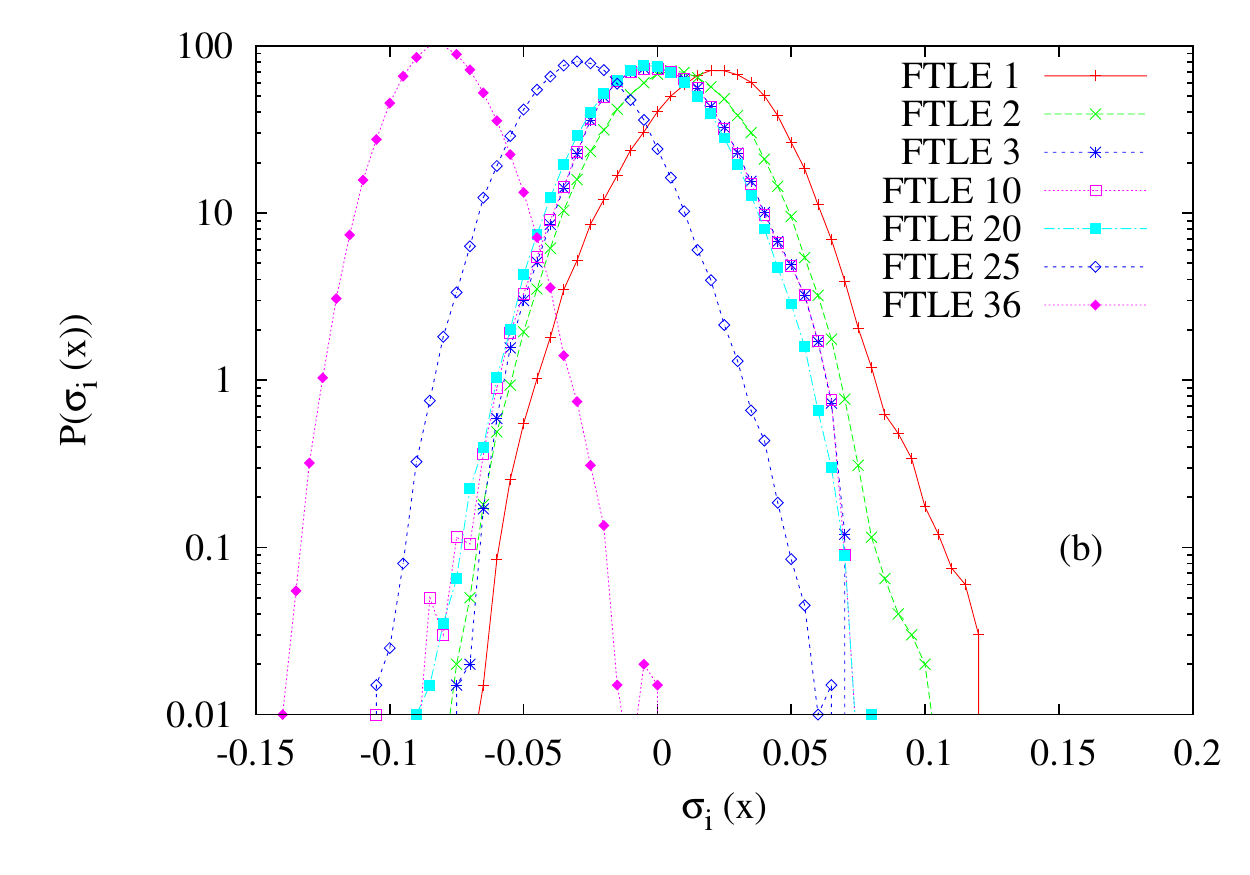}}
\caption{Fluctuations of the instantaneous FTLEs. (a) Time variance of the instantaneous FTCLEs. FTFLEs, and FTBLEs. (b) Probability density of some of the instantaneous FTCLEs. Parameters' value: $C_o=350$ W m$^{-2}$ and $d=1 \times 10^{-8}s^{-1}$.}
\label{distrib_d1x10-8}
\end{figure}

Figure \ref{distrib_d1x10-8}a displays the time variance of the various instantaneous FTLEs. For a given $j$ the time variability of $\sigma_{j,C}^0(k)$, $\sigma_{j,F}^0(k)$, and $\sigma_{j,B}^0(k)$ can be wildly different.  A very interesting feature is that the variance of the local amplification of the covariant vectors is much larger than the ones of the forward and backward vectors for a large group of LEs.   As the CLVs span in a natural way the tangent space, we argue that the variability of the corresponding FTCLEs provides a description of the inhomogeneity of the underlying attractor, as opposed to the variability of the FTFLEs and FTBLEs. 

Figure \ref{distrib_d1x10-8}b displays the probability density of the instantaneous FTCLEs. Interestingly, all of these 
distributions are close to a gaussian shape, with slight asymmetries for the positive and close to 0 exponents, suggesting that the two first moments presented  
in Figs. \ref{Lyap_d1x10-8} and \ref{distrib_d1x10-8}a are already sufficient to describe the inhomogeneity of the attractor.

We wish to extend this analysis by first looking at the time correlations between the time series of $\sigma_{i,X}^0(k)$ and $\sigma_{j,X}^0(k)$, with $i\neq j$, in order to capture the extent of coherence in the growth or decay behavior across the spectrum of exponents, and then by studying the fluctuations of FTLEs constructed as long but finite time averages  of the instantaneous FTLEs. 

\begin{figure}[ht]
\centering
a){\includegraphics[trim=1cm 6cm 1cm 6cm,clip,width=70mm]{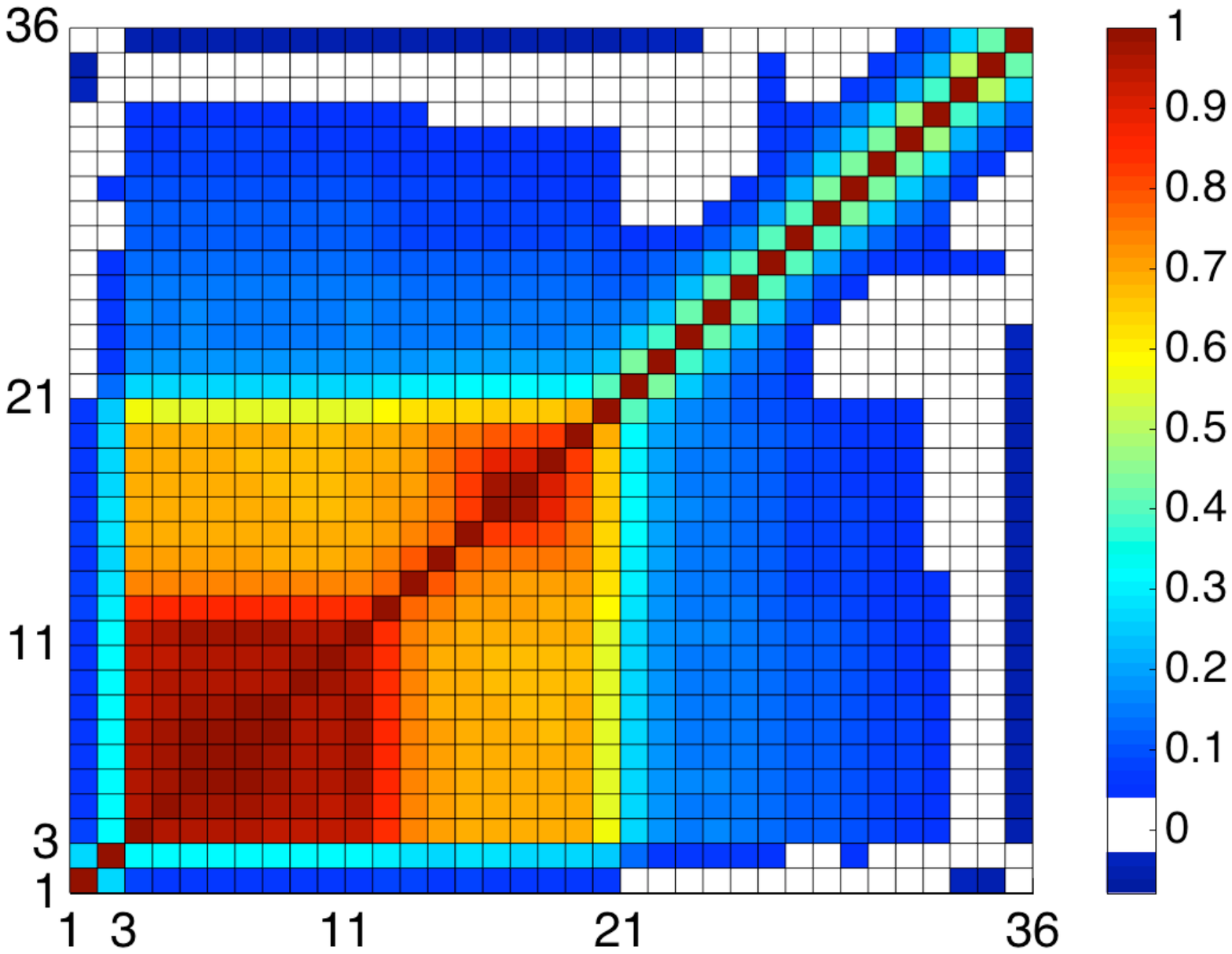}}
b){\includegraphics[trim=1cm 6cm 1cm 6cm,clip,width=70mm]{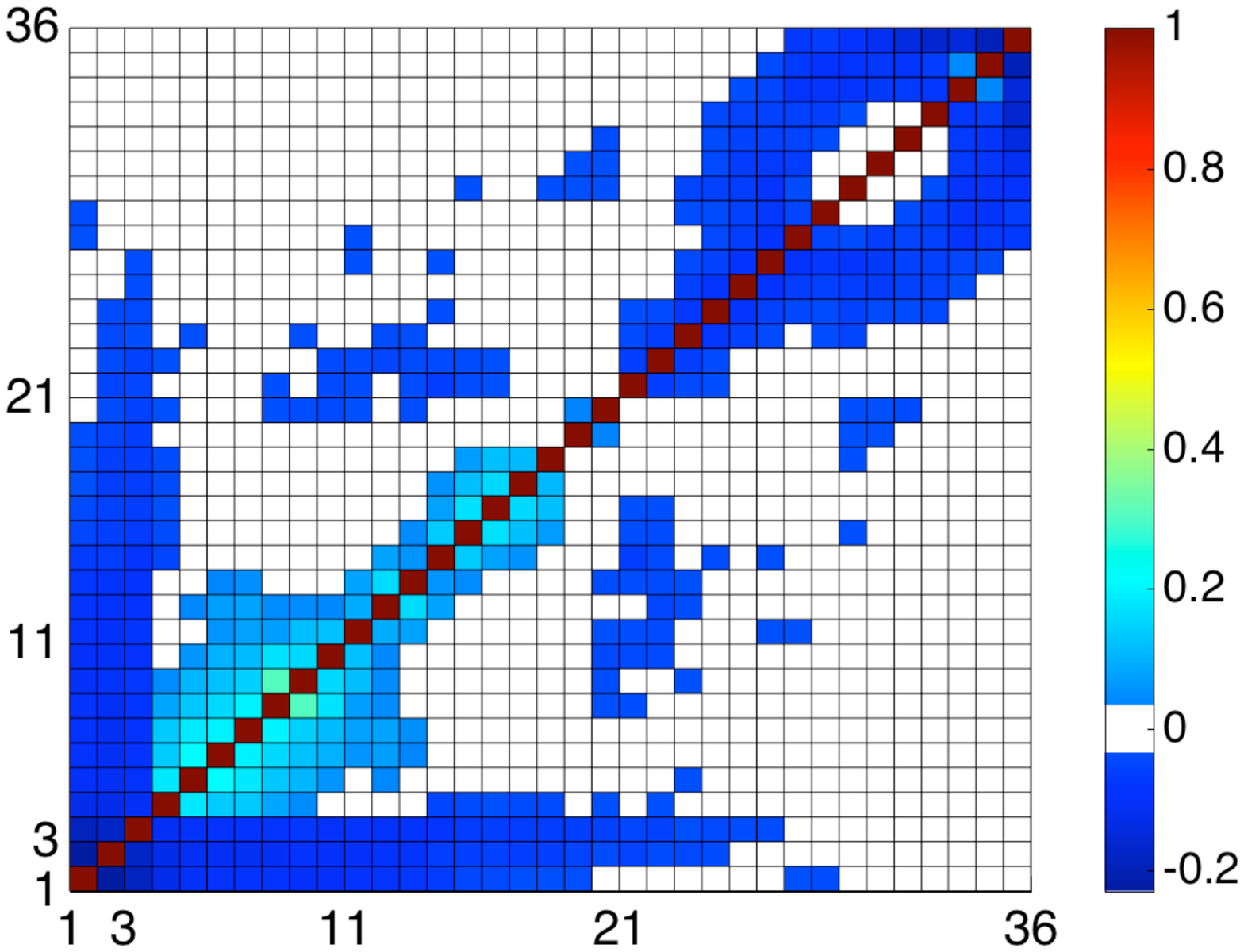}}
c){\includegraphics[trim=1cm 6cm 1cm 6cm,clip,width=70mm]{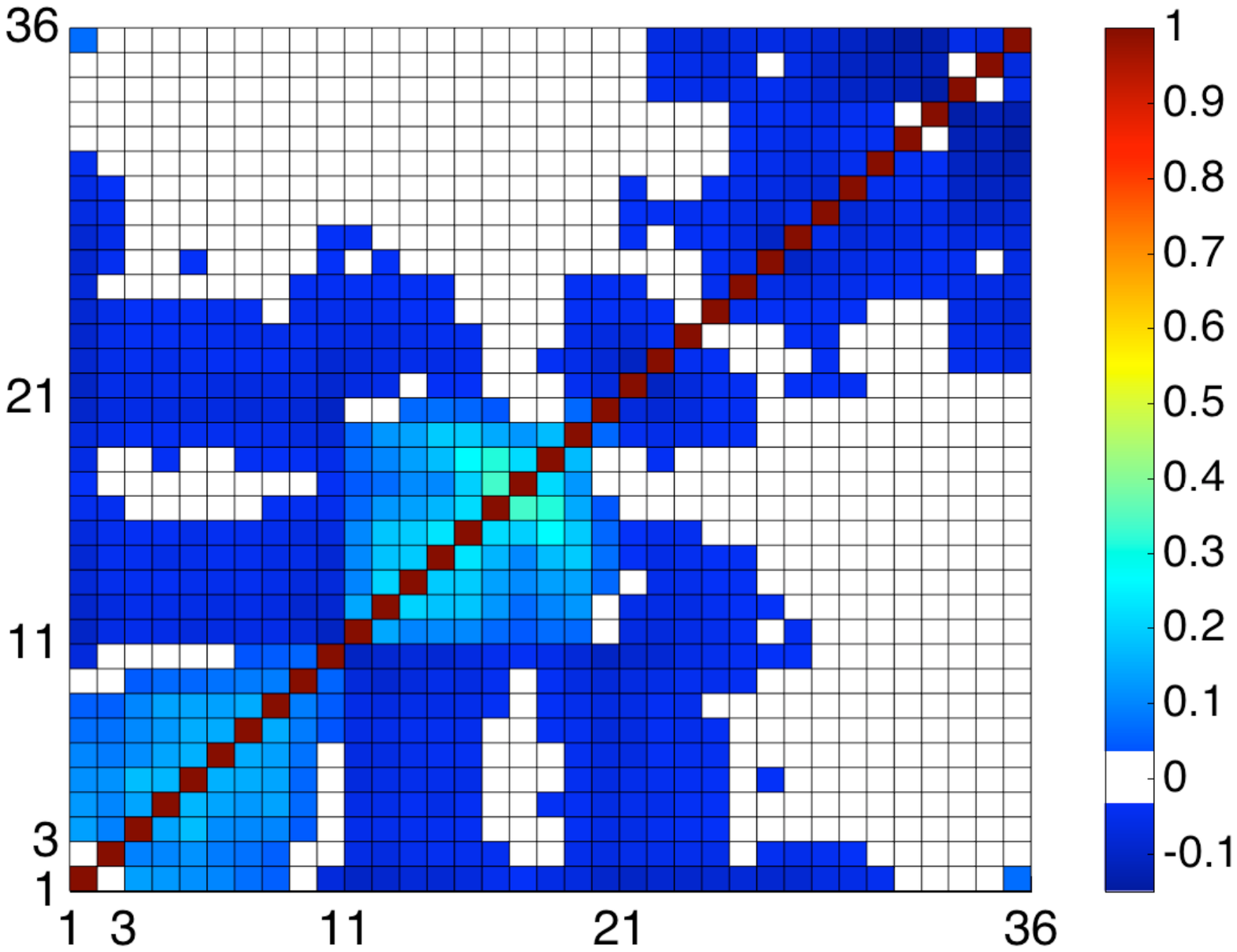}}
\caption{Matrices $\Sigma_{ij}^C$ (panel a) , $\Sigma_{ij}^F$ (panel b), and $\Sigma_{ij}^B$ (panel c) describing the correlation between the time-step estimates of the $i^{th}$ and $j^{th}$ FTCLE, FTFLE, and FTBLE, respectively. See text. Parameters' value: $C_o=350$ W m$^{-2}$ and $d=1 \times 10^{-8}s^{-1}$.}
\label{correlation}
\end{figure}

Figure \ref{correlation} shows the zero-time lag correlation matrix between the time series of $\sigma_{i,C}^0(k)$ and $\sigma_{j,C}^0(k)$ ($\Sigma_{ij}^C$, panel a), $\sigma_{i,F}^0(k)$ and $\sigma_{j,F}^0(k)$, ($\Sigma_{ij}^F$ panel b), and $\sigma_{i,B}^0(k)$ and $\sigma_{j,B}^0(k)$, ($\Sigma_{ij}^B$, panel c), with $i\neq j$. We have indicated with white shading the values that can (roughly) be ruled out as statistically not significant against the null hypothesis of independent series, given the length and the autocorrelation properties of the time series considered here. As immediately apparent, these three matrices substantially differ from each other. This substantiates the fact that while the asymptotic value of the LEs is the same no matter the way they are computed, the statistical properties of the FTLEs can differ substantially.
The properties of $\Sigma_{ij}^C$ stand out as most interesting. {\sv The fluctuations of the instantaneous FTCLEs feature for the most part  positive correlations, with values close to unity for the instantaneous FTCLEs inside the block 3-11 and values larger than 0.7 for the block 12-2. High correlations (values larger than 0.6) are also found between these two blocks.} These results match with the geometrical characterization of the CLVs described previously. The presence of high geometrical degeneracy resulting from many  situations where the CLVs 3-20 are almost exactly parallel to each other is, rather naturally, associated with high correlations in the estimates of the FTLEs. This suggests that the coupling between the perturbations in the CLVs 3-20 will have nontrivial results in terms of error growth and predictability. Additionally, blocks where correlations are extremely high (blocks 3-11 and 12-20) correspond to CLVs having similar distribution of the variance among the system's variables. Finally, one understands that the slow thermodynamic ocean variables are responsible for ensuring the relatively high correlations between the FTCLES of blocks 3-11 and 12-20, because only the former correspond to CLVs prjoecting substantially on the ultraslow dynamic ocean variables.

Outside the region 3-20, the correlations are relatively small, yet in many cases statistically significant. Therefore, there is a tendency of the FTCLEs to fluctuate coherently; this provide an additional interesting characterization of the features of the tangent dynamics.

$\Sigma_{ij}^B$ and $\Sigma_{ij}^F$ are less interesting, as they mostly feature relatively small values, both positive and negative, and in many cases the correlation is not statistically significant. Therefore, the instantaneous FTFLEs and FTBLEs  are by and large independent of each other. Some remnants of the properties of the dynamics described above can be found in  in $\Sigma_{ij}^B$, where relatively high positive correlations can be found inside two blocks, comprising the FTLEs 1-11 and 12-20, respectively.  When considering $\Sigma_{ij}^F$, we find that correlations are small everywhere except for the block of FTLEs 4-12 and for FTLEs of nearby order between 13 and 20.

\subsubsection{Large Deviations Laws (or Lack thereof)}


In order to bridge the gap between the properties of the instantaneous FTLEs and of the asymptotic LEs, we now look into the statistical properties of the FTLEs  $\sigma_{j,X}^M$,  where $X=$ $C$, $F$, and $B$, defined before as the finite M-time average of the value of the instantaneous FTLEs. For well-behaved dynamical systems, such as Axiom A or systems obeying the chaotic hypothesis, one, following \cite{Pazo2013,Schalge2013}, may expect to describe the fluctuations of {\sv $\sigma_{j,X}^M$} for $M\gg1$ but finite using a large deviations law \cite{Kifer90}.  Large deviations theory deals with the study of the statistical properties of the averages of many but finite stochastic variables (typically also identically distributed) and is the backbone of modern statistical mechanics (see, \textit{e.g,} \cite{Touchette09}). In this case, the goal is to discover whether we can write:   
\begin{equation}
\mathcal{P}(\sigma_{j,X}^M=x)\sim\exp[-MI^X_j(x)]
\end{equation} 
where $I^X_j(x)$ is the so-called rate function, which has a minimum for $x={\sigma}_j$ $\forall X$.The previous expression indicates an exponential convergence of the estimates of the FTLEs to the asymptotic value as the averaging time is increased. 

Note that  it is relatively easy to derive, under suitable conditions, the rate function when the stochastic variables that are averaged are independent and identically distributed. In this case, one can combine the Cramer theorem and the Gaertner-Ellis theorem {\sv as follows. One first introduces the} cumulant generating function $$\lambda(k)=\log \int \mathrm{d}x \psi(x) \exp[kx],$$ where $\psi(x)$ is the probability density function (pdf) of each stochastic variable \cite{Touchette09}. {\sv Then the rate function $I(x)$ is derived as a Legendre transform of $\lambda(k)$, so that $$I(x) =  \sup_{k\in R}\{kx-\lambda(k)\}$$} 
In our case the condition of independence is clearly broken because for each $j$ the time series $\sigma_j^k)$, $k=1,\ldots$ feature a nontrivial autocorrelation. The easiest way to proceed for deriving the rate functions in presence of nontrivial autocorrelation properties, as in the case of a chaotic dynamical system, is to just construct empirically the pdf $\mathcal{P}(\sigma_{j,X}^M=x)$ and test whether the limit $\lim_{M\rightarrow \infty} [-\log(\mathcal{P}(\sigma_{j,X}^M=x)) /M ]$ exists. If this is the case, the limit gives the rate function of interest.  

We anticipate that in this work, we do not have the goal of a quantitative characterization of the rate functions, but we would rather stick to a discussion of their qualitative properties. An accurate quantitative study of the convergence - beyond checking graphically that all curves collapse to a universal limit curve - would require very long integrations. Still, our results are possibly quite useful. We report in Table \ref{mytable}  a qualitative description of the rate functions describing the large deviation laws of the exponents of the finite time Covariant (FTCLE), Forward (FTFLE), and Backward (FTBLE) Lyapunov Vectors. The $j^{th}$ row of the table describe the properties of the $j^{th}$ FTCLE, FTFLE, and FTBLE. In each entry of the table we report a capital letter followed by a small letter, in the form $Xy$. The first letter indicates whether the rate function grows asymptotically linearly ($X=L$) or with higher power ($X=Q$). If {\sv for a given value of $j$} two {\sv or more} rate functions are identical, they share the same  pair of letters. Instead, if they have qualitatively similar but quantitatively different asymptotic behaviour, then the second letter differs.

\begin{table}[]
\centering
\caption{Qualitative properties of the rate functions characterizing the large deviations of the exponents of the finite time Covariant (FTCLE), Forward (FTFLE), and Backward (FTBLE) Lyapunov Vectors. With the first capital letter we indicate whether the rate function is asymptotically linear ($L$), quadratic or with higher power ($Q$), or no rate function can be detected (X). {\sv If the first and second letter in the two or more entries in a given row coincide, then the corresponding rate functions are identical}. Note that the rate functions are identical for the FTCFEs, FTFLEs, and FTBLEs corresponding to the non-zero LEs (order 1,2, 21-36). Parameters' value: $C_o=350$ W m$^{-2}$ and $d=1 \times 10^{-8}s^{-1}$.}
\label{mytable}
\begin{tabular}{@{}llllllll@{}}
Order & FTCLE & FTFLE & FTBLE & Order & FTCLE & FTFLE & FTBLE \\
1     & Qa  & Qa  & Qa  & 2     & Qa & Qa  & Qa \\
3     & La  & Lb  & Lc  & 4     & La  & La  & Lc   \\
5     & La  & Lb  & Lc  &  6     & La  & Lb  & Lc   \\
7     & La  & Lb  & Lc  &  8     & La  & Lb  & Lc   \\
9     & La  & Lb  & Lc  &  10    & La  & Lb  & Lc  \\
11    & Qa  & Qa  & Xa  &  12    & Qa  & Qa  & Xa   \\
13    & Qa  & Qb  & Xa  &  14    & Qa   & La  & Lb  \\
15    & Qa  & La  & Lb  & 16    & Qa   & La  & Lb  \\
17    & Qa   & La  & Lb &  18    & Qa   & La  & Lb  \\
19    & Qa   & La  & Lb &  20    & Qa   & La  & Lb  \\
21    & Qa   & Qa  & Qa &  22    & Qa   & Qa  & Qa  \\
23    & Qa   & Qa  & Qa &  24    & Qa   & Qa  & Qa  \\
25    & Qa   & Qa  & Qa &  26    & Qa   & Qa  & Qa  \\
27    & Qa   & Qa  & Qa &  28    & Qa   & Qa  & Qa  \\
29    & Qa   & Qa  & Qa &  30    & Qa   & Qa  & Qa  \\
31    & Qa   & Qa  & Qa &  32    & Qa   & Qa  & Qa  \\
33    & Qa   & Qa  & Qa &  34    & Qa   & Qa  & Qa  \\
35    & Qa   & Qa  & Qa &  36    & Qa   & Qa  & Qa  \\
\end{tabular}
\end{table}

We have found  that the rate functions can be clearly defined for $j=1,2$ and $j=21-36$, \textit{i.e.} for the nonzero LEs. The fast decay of correlation for these FTLEs contributes to this result. Additionally, we have that $I^C_j(x)=I^F_j(x)=I^B_j(x)$, thus indicating that the various methods for constructing LEs provide approximately identical (apart from logarithmic corrections) statistics in such an asymptotic regime.

Such universality contrasts with what seen in Figs. \ref{distrib_d1x10-8}a and Figs. \ref{correlation}a, namely that  the fluctuations of the FTCLEs, FTFLEs, and FTBLEs at the time step are in general  different, both in terms of variance of each FTLEs and correlation between the different FTLEs. Note that there is no general theoretical argument supporting the identity between  $I^C_j(x)$, $I^F_j(x)$, and $I^B_j(x)$, as only the asymptotic values of the LEs are well-known to be the same no matter whether {\sva they are computed} following the covariant, forward, and backward Lyapunov vectors. See discussion in \cite{Laffargue13} and \cite{Politi15}. Note that, taking a gaussian approximation, and so more in the spirit of the central limit theorem than of actual large deviations, a high degree of universality is obtained for the fluctuations of the long-term averages of the FTLEs \cite{Pazo2013}.

For FTLEs 1-2 and 21-36, the rate functions grow asymptotically with at least a second power, and in some cases lack of symmetry around the minimum is apparent. Importantly, the rate function of the first and second FTLE crosses the zero, and so do some of the negative FTLEs with index larger than 21 {\sv (not shown)}, thus suggesting the presence of rare but existing tangencies between the unstable and stable manifold, which indicates the lack of uniform hyperbolicity. Nonetheless, as hinted at before, in this system deviations from hyperbolic behaviour come essentially from the presence of many quasi-zero LEs. 

The complex nature of the dynamics of the system corresponding to the quasi-zero LEs emerges again {\sv in the results relative to the FTLEs 3-20.} In this case, the convergence of $-\log(\mathcal{P}(\sigma_{j,X}^M=x)) /M$ to the rate function is {\sv systematically} slower {\sv for FTLEs 3-10}, and {\sv in some cases, for FTLEs 11-20.} {\sv Most importantly,} there is no agreement between $I^C_j(x)$, $I^F_j(x)$, and $I^B_j(x)$. The differences can be both quantitative and qualitative: for $j=3-10$ the three rate functions are asymptotically linear, but disagree on the linear coefficients, while for, \textit{e.g}, $j=14-20$ we have that $I_j^C(x)$ grows asymptotically with at least a quadratic power, while $I_j^F(x)$ and $I_j^B(x)$ are asymptotically linear (but quantitatively different).

\section{Dynamics of the error}
\label{sec:error}

The fluctuations in time of the local amplifications  have strong implications on the error dynamics as discussed in details in \cite{Benzi1989} and in
 \cite{Nicolis1995}. In the latter, in particular, a discussion of the impact of the norm on the dynamics of small finite amplitude errors of highly inhomogeneous 
systems indicates that a strong superexponential growth (faster than the amplification associated with the first LE) is experienced when the
classical $L^2$ norm is used. 
In the previous section, the variability along the CLVs has been investigated in details and a complicate picture emerged, with a variability
along a set of CLVs ([1, 2] and [21-36]) compatible with the behavior identified in the classical large deviations theory \cite{Kifer90,Touchette09}, and a set [3-20]
for which the convergence (if any) toward their asymptotic values is very slow. This also suggests that local fluctuations of the amplifications
of small errors in this subspace can be highly non-trivial and could lead to complicate predictability properties. This aspect is investigated
in the following.

Before discussing this in details, let us first recall how the predictability of a system is usually evaluated in an operational 
context of weather and climate forecasts.  
Let us consider a solution of the system in phase space, $\vec{x}(t_0)=\vec{x}_0$, at time $t_0$. Observations of this system are affected by finite-amplitude 
initial errors that can be 
for simplicity be considered as a gaussian white noise, $\epsilon(x_0)$, and the observed state is then, $\vec{x}'(t_0)=\vec{x}_0+\vec{\epsilon}(x_0)$.   
One can now measure the error evolution starting from these two initial conditions as
\begin{equation}
\vec{E}(t)=\vec{x}'(t)-\vec{x}(t)
\end{equation}
where $\vec{x}'(t)$ and $\vec{x}(t)$ are the two trajectories starting from the two initial conditions $\vec{x}'(t_0)$ and $\vec{x}(t_0)$ of the perturbed and
unperturbed trajectories. Since the amplification of this error is fluctuating on the inhomogeneous attractor of the system, already discussed in the previous section, an average
over the attractor is necessary in order to get properties that are independent of the initial state. The classical norm used is the $L^2$ norm as defined 
\begin{equation}
\langle E_t^2\rangle = \int \mathrm{d}\vec{\epsilon}_0 \rho_{\epsilon} (\vec{\epsilon}_0) \int \mu(\mathrm{d}\vec{x}_0) \left[ (\vec{x}'(t) - \vec{x}(t)) \cdot (\vec{x}'(t) - \vec{x}(t)) \right] 
\label{L2norm}
\end{equation}
where $\rho_{\epsilon}(\vec{\epsilon}_0) \mathrm{d}\vec{\epsilon}_0$ and $\mu(\mathrm{d} \vec{x}_0)$, are the probability measure of the initial errors and of the initial conditions
on the attractor of the system. The amplitude of the perturbations $\vec{\epsilon}$ is taken
sufficiently small in order to get information of the full error growth evolution, namely the 
exponential, linear and saturation regimes, see \cite{Nicolis1995, Vannitsem1994}.  

Another particular norm is the logarithmic norm defined as 
\begin{equation}
\langle \ln E_t^2 \rangle = \int \mathrm{d}\vec{\epsilon}_0 \rho_{\epsilon} (\vec{\epsilon}_0) \int \mu(\mathrm{d}\vec{x}_0) \ln \left[ (\vec{x}'(t) - \vec{x}(t))  \cdot (\vec{x}'(t)) - \vec{x}(t)) \right]
\label{expnorm0}
\end{equation}
which is leading to a very simple expression of the error evolution when the perturbations are aligned along the CLVs,
\begin{equation}
\left \langle \ln \frac{E_t^2}{ \epsilon_{i,0}^2} \right \rangle = 2 \sigma_i t 
\label{expnorm1}
\end{equation}
for $\epsilon_{i,0}$ a sufficiently small perturbation along the $i$th covariant vector (and for short times).
 
As discussed in \cite{Nicolis1995}, the use of the $L^2$ norm is strongly affected by the inhomogeneity of the attractor - for small initial error amplitudes and for
short times - because the average of an exponential amplification of the error as arising from the use of Eq. (\ref{L2norm}) is not equal to the exponential of 
the average as defined by (\ref{expnorm1}). In addition, if the initial error is not aligned along the covariant vectors, both the $L^2$ norm and the logarithmic
norm will be affected by the inhomogeneity of the attractor, \textit{i.e.} the variability of the local stretching rates. The inhomogeneity of the attractor 
mostly affects the $L^2$ norm by inducing a superexponential behavior (a growth rate faster than the one associated with the LE) with a rate proportional to
the variance of the local stretching factors. For the logarithmic norm, the inhomogeneity could also show up in the error dynamics provided that the initial error 
is not aligned along the covariant vectors.     

In order to illustrate this dynamics, one can again refer to the detailed analysis of \cite{Nicolis1995} who showed that when
 random perturbations are introduced in a 2 dimensional subspace in which the dynamics is developping (e.g. spanned by two CLVs), the 
error evolution measured by these two norms are,
\begin{align}
\langle E(t)^2\rangle& = \langle\epsilon_1^2\rangle \exp(2 \sigma_1 t) \left\langle\exp\left(2 \int_0^t  \mathrm{d}\tau S_1(\tau) d\tau\right)\right\rangle \\ 
&+ \langle\epsilon_2^2\rangle \exp(2 \sigma_2 t) \left\langle \exp\left(2 \int_0^t S_2(\tau)\right)\right\rangle
\end{align}
where $\epsilon_i$ and $S_i(t)=\sigma^0_i-\sigma_i$ are the initial random perturbation along direction i and the fluctuations of the LFTEs along the ith CLV. 
For the logarithmic norm, one finds 
\begin{eqnarray}
\left\langle\ln \frac{E_t^2}{\epsilon_{i,0}^2} \right\rangle = 2 \sigma_i t +\nonumber \\ 
\left < \ln \left ( 1 + \frac{\epsilon_2^2}{ \epsilon_1^2+\epsilon_2^2} \left (exp \left [ 2 (\sigma_1-\sigma_2) t + 2 \int_0^t \mathrm{d}\tau (S_2(\tau) - S_1(\tau)) \right ] -1 \right ) \right ) \right > 
\end{eqnarray}

\begin{figure}[ht]
\centering
a){\includegraphics[width=70mm]{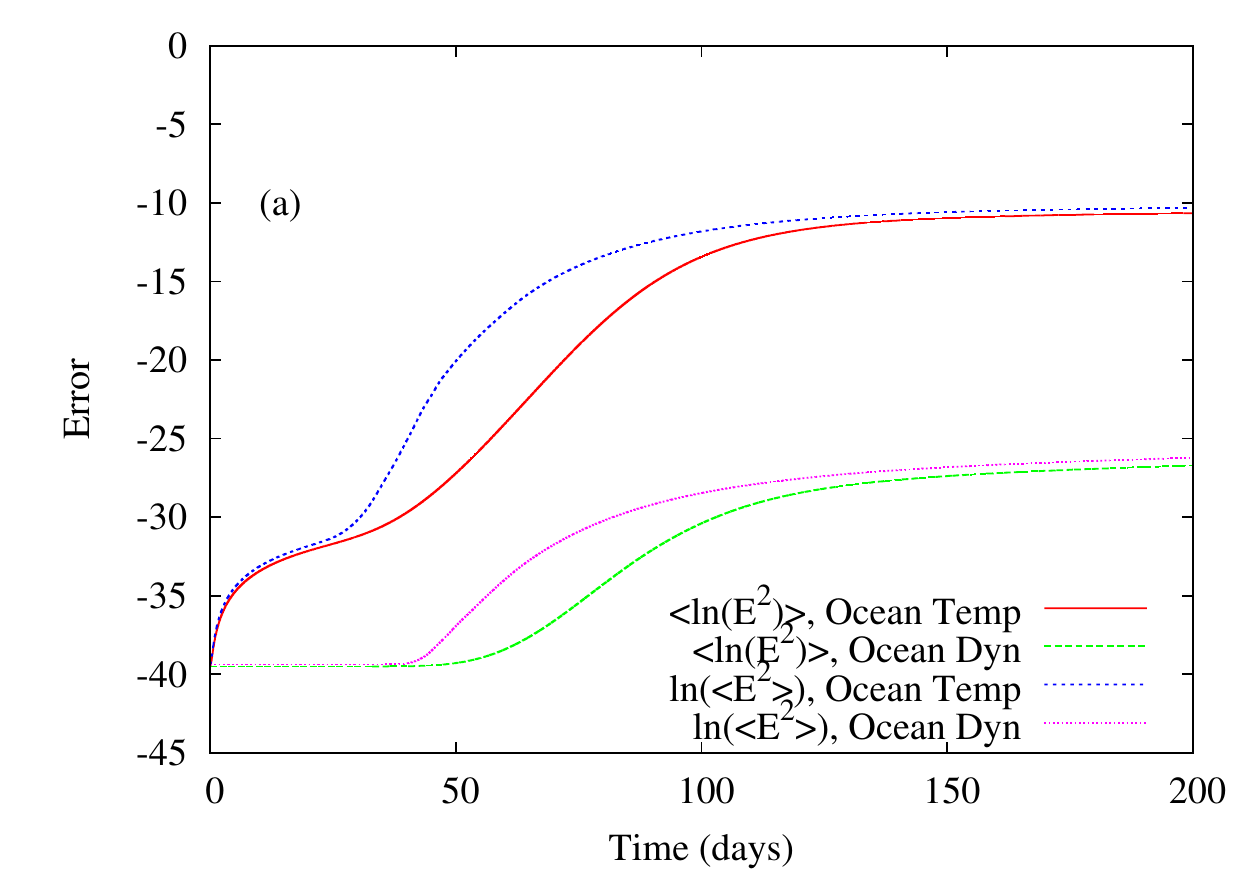}}
b){\includegraphics[width=70mm]{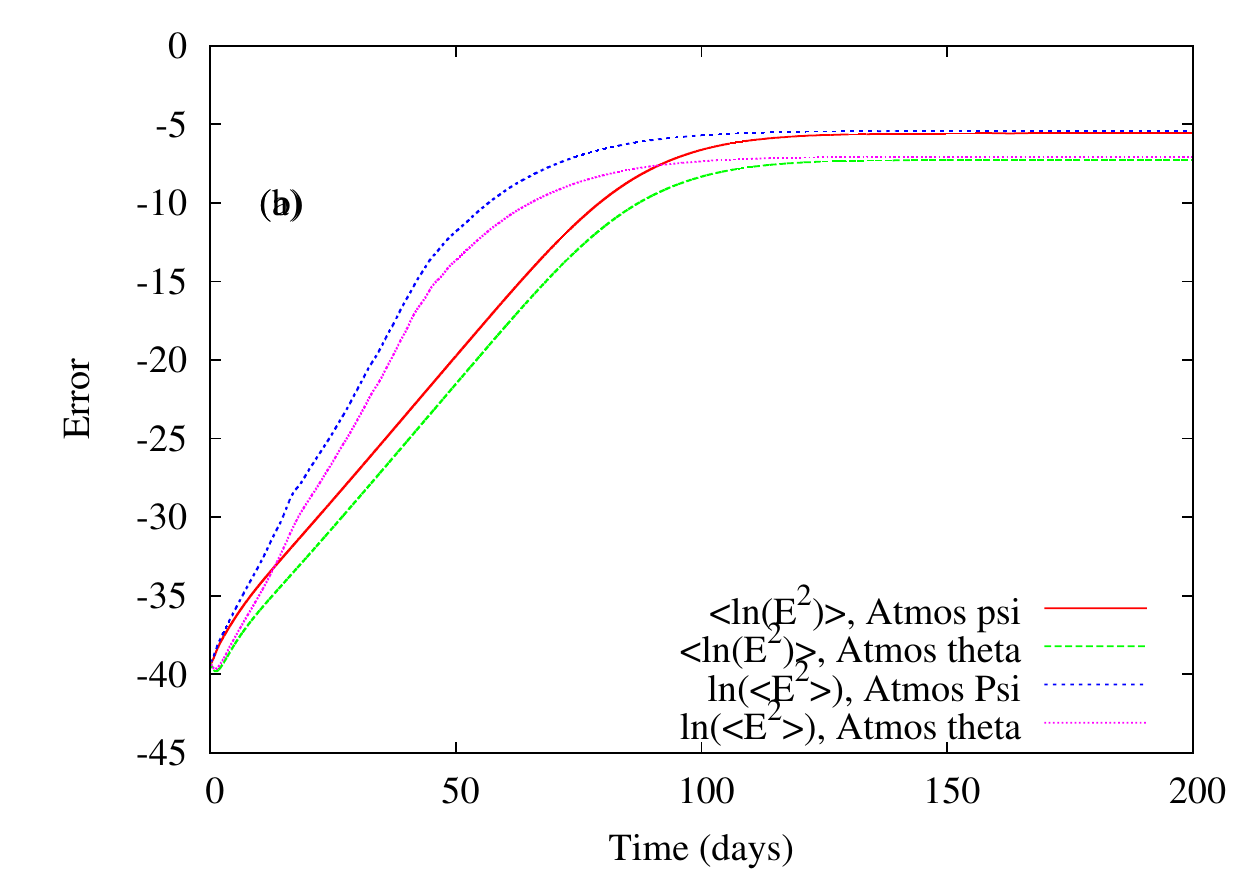}}
c){\includegraphics[width=70mm]{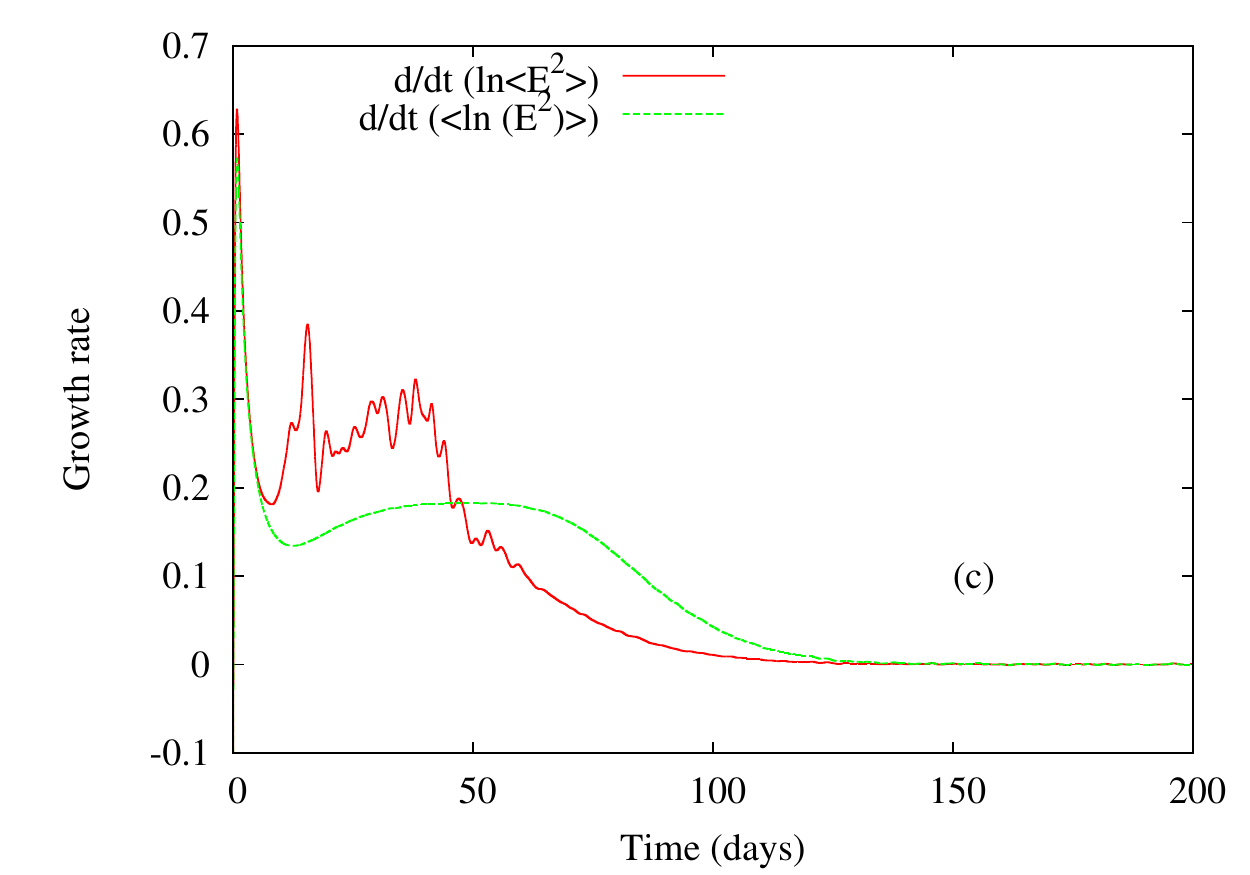}}
\caption{Evolution of the averaged norm of the error as obtained from an averaged over 100,000 realizations. Two norms are used (and referred in the 
legend of each panel), the $L^2$ norm and the logarithmic norm. (a) Evolution based on random initial perturbation. The four curves correspond to
the logarithmic norm for the ocean temperature (continuous red line), for the ocean transport (dotted green line), the $L^2$ norm for the ocean
temperature (blue dotted line) and the $L^2$ norm for the ocean transport (black dotted curve). (b) as in (a) but for the atmospheric variables 
{\sv, the barotropic streamfunction, $(\psi^1_a + \psi^3_a)/2  $, (referred to 'Atmos psi' in the panel) and the baroclinic streamfunction, $(\psi^1_a - \psi^3_a)/2 $, {\sva also related to the temperature field as shown at sec. \ref{ssec:ocean_temps} }(referred to 'Atmos theta' in the panel)}. (c) Growth rate for the total error (averaged over all the variables) for the $L^2$ norm (red curve) and the logarithmic norm (green curve). Parameters' value: $C_o=350$ W m$^{-2}$ and $d=1 \times 10^{-8}s^{-1}$.}
\label{Fig:randerr}
\end{figure}
For the $L^2$ norm, the superexponential growth is associated with the amplitude of the fluctuations, as discussed in detail in \cite{Benzi1989, Nicolis1995}, whichever the orientation of the perturbation is. For the logarithmic norm, the fluctuations will affect the dynamics only when the perturbations
are not aligned along one specific CLV, whose impact can be either subexponential or superexponential depending on the statistical and dynamical
properties of the fluctuations \cite{Nicolis1995}. 

\begin{figure}[ht]
\centering
a){\includegraphics[width=70mm]{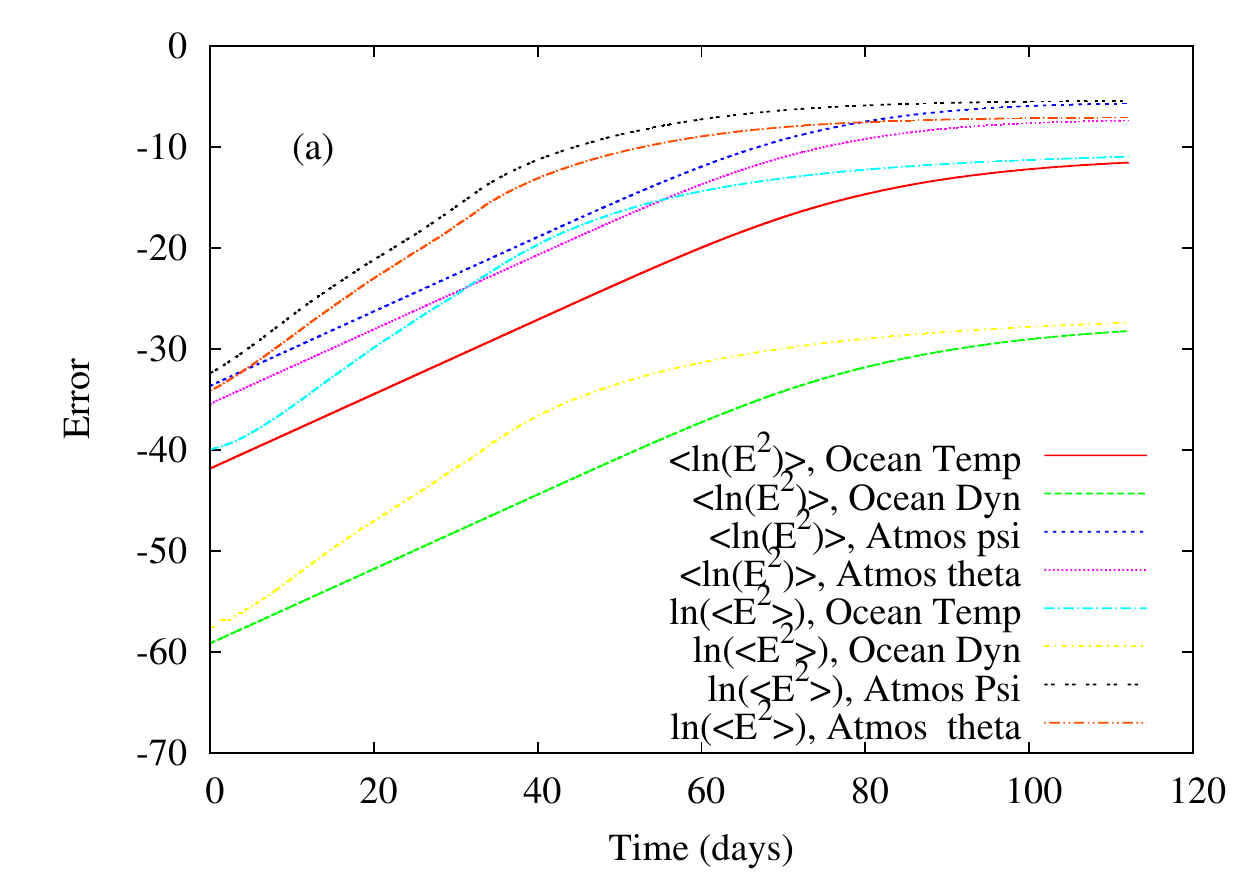}}
b){\includegraphics[width=70mm]{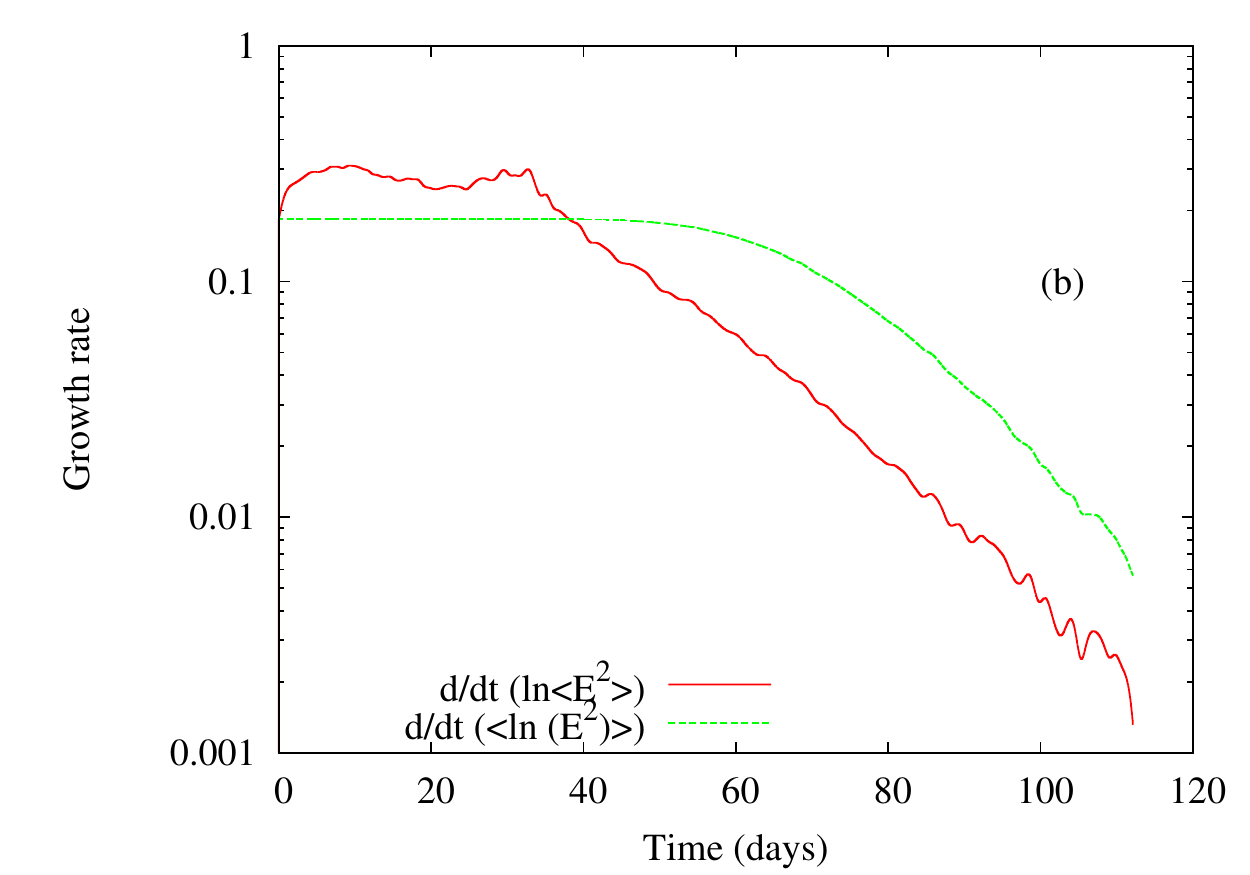}}
c){\includegraphics[width=70mm]{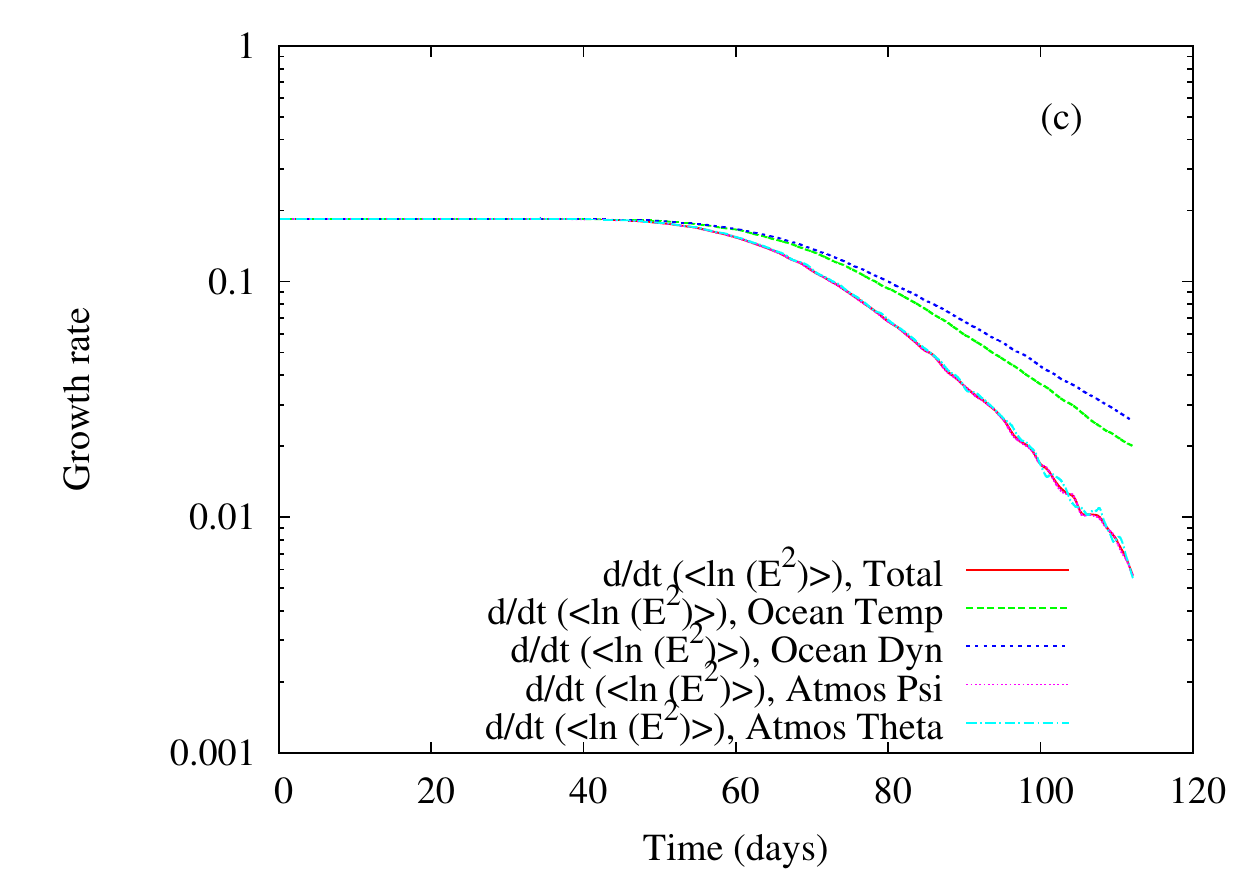}}
\caption{(a) Curves as in panels (a) and (b) of Fig. 6, but for a perturbation along the 1st CLV. (b) Growth rate along the firs CLV for the $L^2$ 
norm  (red curve) and the logarithmic norm (green curve). (c) Growth rates for the logarithmic norm for the different fields of the system,
namely the total error growth (red line), the  ocean temperature (green line), the ocean transport (blue line), the atmospheric barotropic streamfunction (magenta line) and the baroclinic streamfunction (black line). Parameters' value: $C_o=350$ W m$^{-2}$ and $d=1 \times 10^{-8}s^{-1}$.}
\label{Fig:randvect1}
\end{figure}

Let us now consider the error dynamics in the coupled ocean atmosphere system, and let us start with a random perturbation in phase space affecting all covariant 
vectors. The perturbation is gaussian with a 0 mean and a standard deviation of $10^{-9}$ in each variable. Figures (\ref{Fig:randerr}a)-(\ref{Fig:randerr}b)       
display the error evolution as a function of time as obtained with 100,000 realizations for the ocean and the atmosphere components, respectively.
In order to compare both the $L^2$ and logarithmic norms, the logarithm of the $L^2$ norm is represented on both panels.

A first interesting feature of the error dynamics is the transient rotation of the error toward the first CLV, associated with a slow
decrease of the error along the ocean streamfunction modes and a rapid amplification along the ocean temperature modes lasting for a period which depends on the
typical time scales associated with each variable (Fig. (\ref{Fig:randerr}a)). Within the atmosphere, the convergence toward the dominant direction of instability
is faster and similar for both the streamfunction and temperature fields. After this rotation, the error displays a dynamics related to the dominant Lyapunov
vector. This rotation impacts both error norms in a quite similar manner. After this early phase drastic differences emerge, with a seemingly   
more rapid amplification of the error for the $L^2$ norm. As alluded before, this
behaviour reflects the inhomogeneity of the underlying attractor, which manifests with different mean error growth rates
for the two norms investigated here. Finally the error evolution experiences a saturation phase during which the linearized hypothesis of the error dynamics 
cannot be made anymore.

This dynamics can be further clarified by computing the effective growth rates as $1/2 \, \mathrm{d}/\mathrm{d}t (\ln \langle E^2_t\rangle)$ and $1/2 \, \mathrm{d}/\mathrm{d}t (\langle \ln E^2_t\rangle)$ displayed in panel (c) of
Fig. (\ref{Fig:randerr}). {\sv If this growth rate is larger than the value of the dominant Lyapunov exponent, then the error growth is 
qualified as superexponential.} 
For the $L^2$ norm, the growth rate is always larger than the value of the dominant LE whose value is $0.181$ days$^{-1}$, {\sv up to a lead time of 45 days}. After {\sv this} superexponential growth, the
error growth rate decreases rapidly in an exponential way. For the logarithmic norm, the growth rate reaches the value of the dominant LE after
about 30-40 days and becomes constant for about 30 days before decreasing when the mean error starts to saturate. These results clearly indicate that the dominant
LE is not a correct measure of the error amplification rate when the $L^2$ norm is used.     

Let us now consider perturbations applied along specific CLVs. First consider a random perturbation aligned along the first covariant vector
of the coupled system as illustrated in Fig. (\ref{Fig:randvect1}). A picture similar to the one presented in Fig. (\ref{Fig:randerr}) can be drawn, except that
the transient rotation toward the first covariant vector has disappeared. Still the strong superexponential growth rate in the $L^2$ norm is present as displayed
in panel (b) of Fig. (\ref{Fig:randvect1}). This also suggests that this superexponential growth rate in intrinsic to the dynamics and not an artefact
due to a potential transient effect. Finally in panel (c), the growth rate for the logarithmic norm is computed separately for the different fields of the 
different components of the system.  Not surprisingly the initial growth rate is the same, but for long lead times when the error starts to saturate, a differential
behavior is observed. The saturation is faster for the atmospheric fields, while it is slower for the ocean temperature and even slower for the ocean streamfunction
field. This obviously suggests that certain variables of the system display a longer predictability in the nonlinear phase of the error growth. A similar picture
at long lead time is found with the $L^2$ norm.    

\begin{figure}[ht]
\centering
a){\includegraphics[width=70mm]{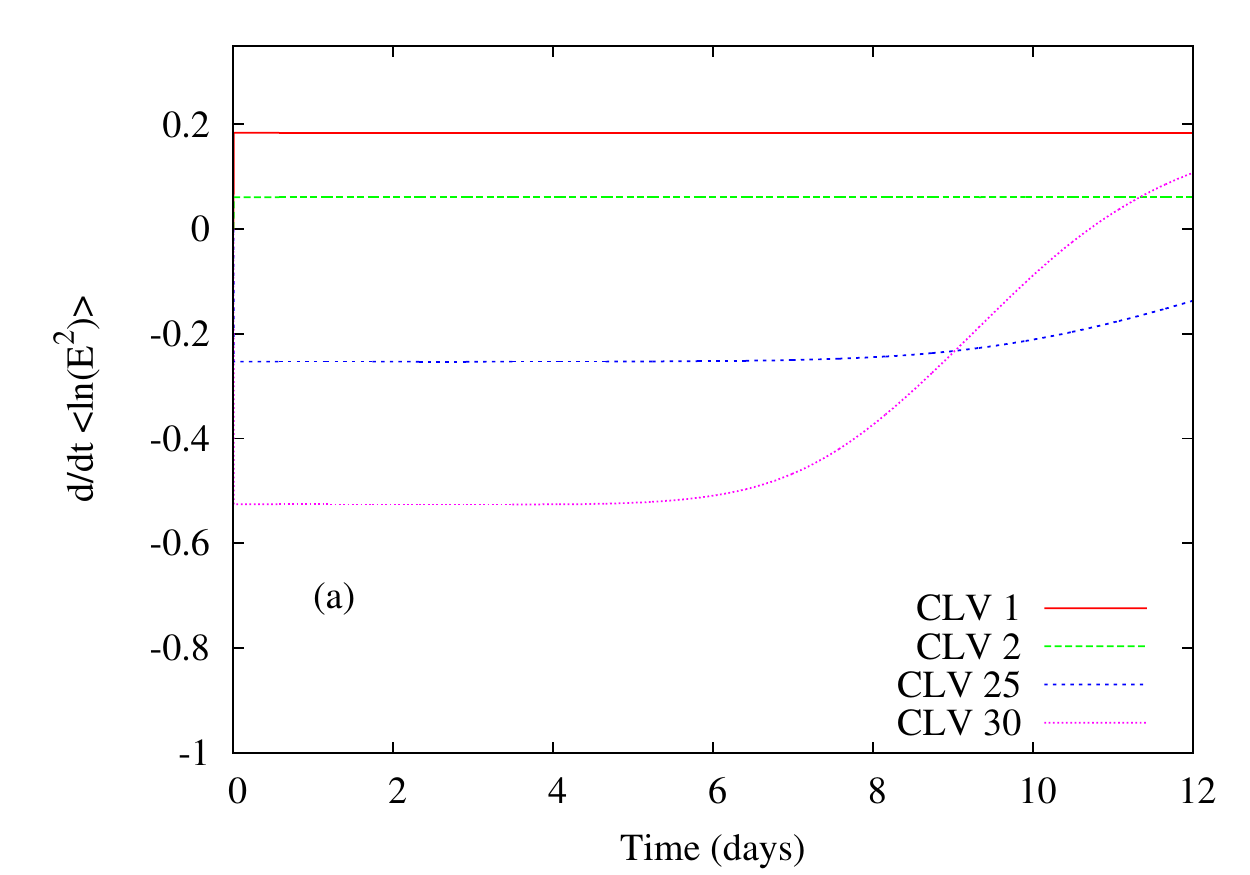}}
b){\includegraphics[width=70mm]{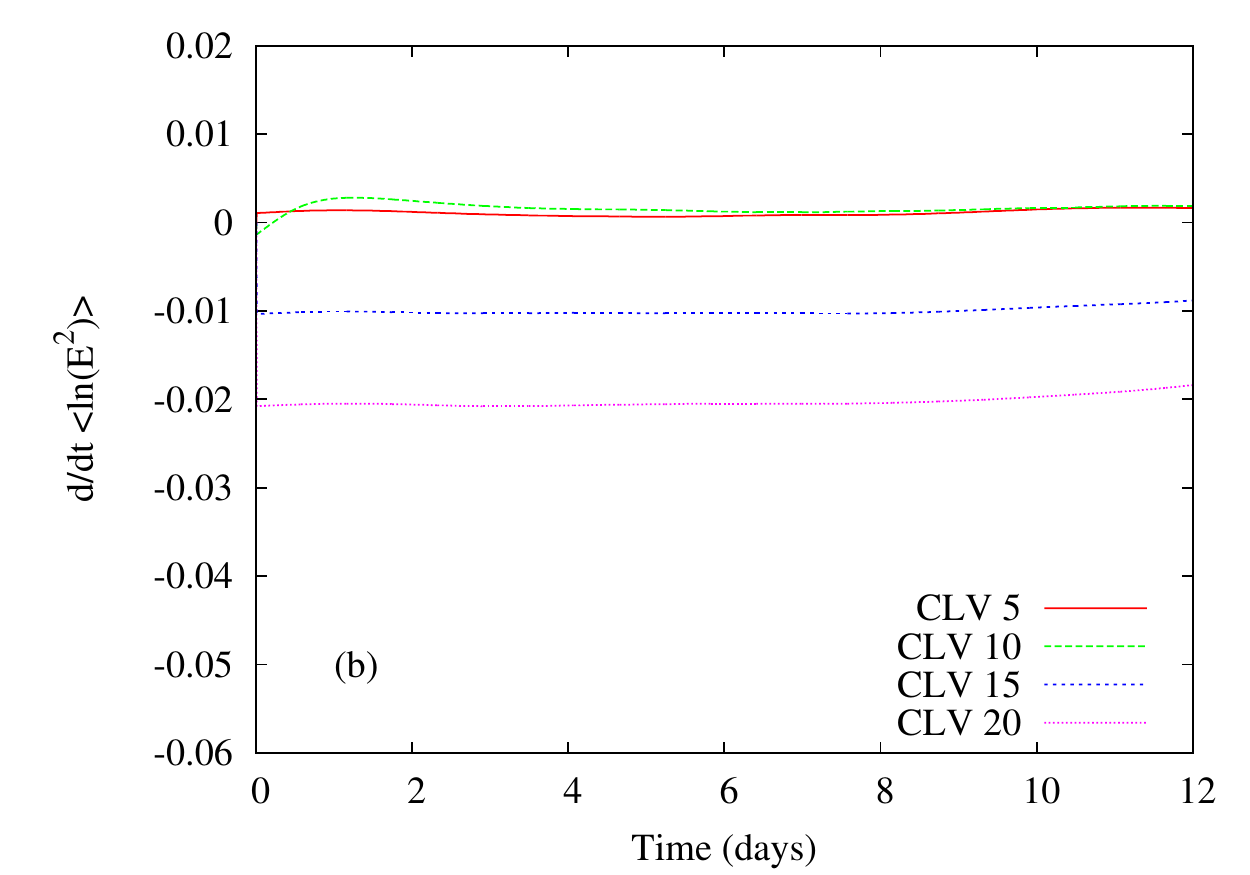}}
c){\includegraphics[width=70mm]{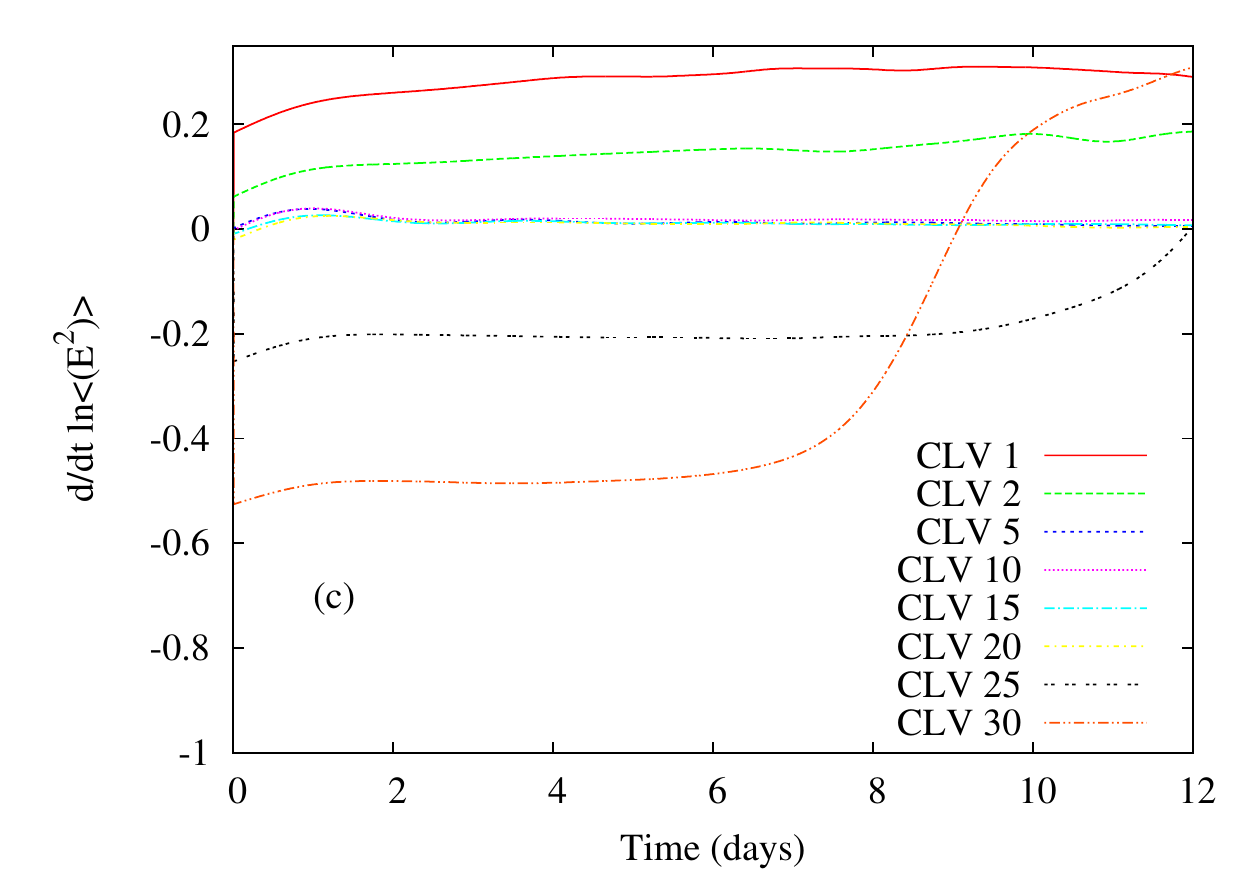}}		
\caption{ Growth rates of the error along different CLVs for the logarithmic norm (a) and (b), and (c) the $L^2$ norm. Parameters' value: $C_o=350$ W m$^{-2}$ and $d=1 \times 10^{-8}s^{-1}$.}
\label{Fig:randvecti}
\end{figure}

Let us now investigate the error dynamics when perturbations are randomly introduced along the other CLVs. 
Figure (\ref{Fig:randvecti}) displays the growth rate of the error evolution averaged over 100,000 realizations for covariant vectors $i=1, 2, 5, 10, 15, 20, 25$ and $30$.
Panels (a) and (b) correspond to the average of the logarithm of the error, $1/2 \, \mathrm{d}/\mathrm{d}t \langle \ln (E^2_t)\rangle$, for two subsets of vectors spanning the unstable space and the center direction, respectively, and (c) the logarithm of the averaged error, 
$1/2 \, \mathrm{d}/\mathrm{d}t \ln\langle(E^2_t)\rangle$. 

The growth rate for the averaged of the logarithm of the error clearly corresponds to the associated LE for short times, 
as expected (further confirming the correct computation of the CLVs), except for CLVs 5 and 10 still displaying a superexponential dynamics even if the corresponding exponents are slightly negative (panel (b)). This interesting feature suggests that these vectors have not converged (yet), if any, toward the corresponding true CLVs.

For long times, it starts to deviate from the perfect exponential behavior due to the presence of round-off numerical errors {\sv -- and the impact
of nonlinearities even if these are small-- as clearly illustrated by the behavior of the error along CLVs 25 and 30 in Fig. 8}. 

Interestingly, the superexponential behavior for the $L^2$ norm averaged error is present whatever the covariant vector considered indicating that
the fluctuations affect all directions in phase space. Additionally, the dynamics of the averaged error for exponents close to 0 displays    
a positive growth rate contrasting with the negative value of the corresponding exponent.

\begin{figure}[ht]
\centering
a){\includegraphics[width=70mm]{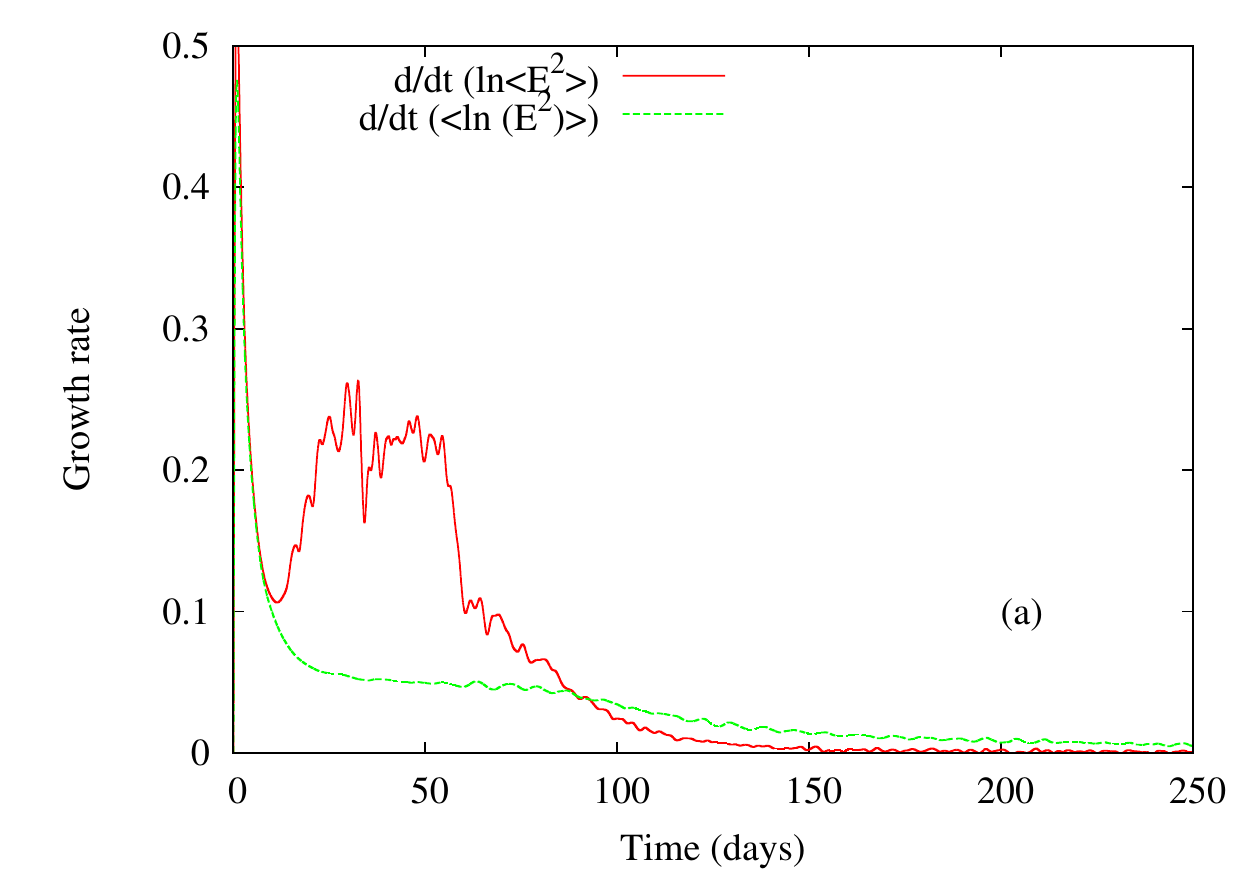}}
b){\includegraphics[width=70mm]{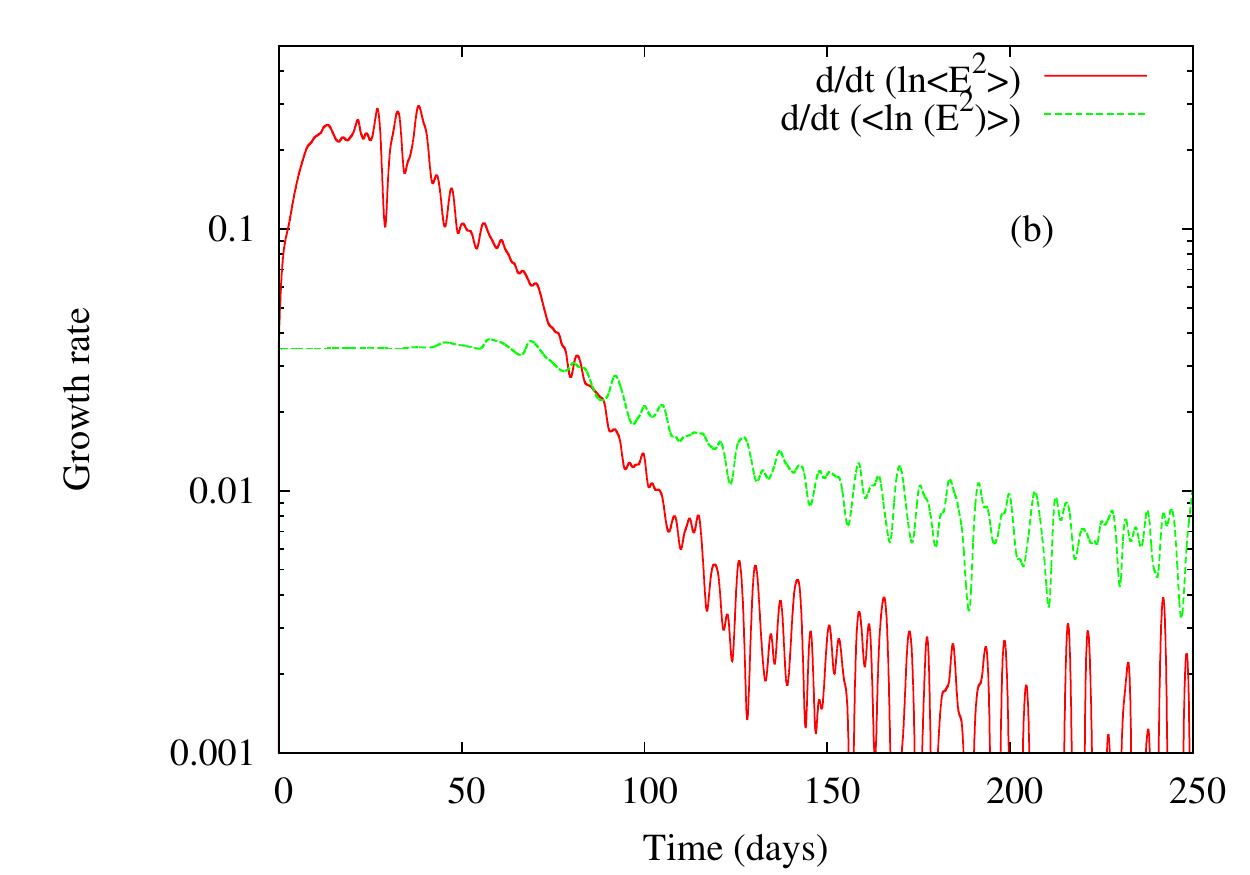}}
c){\includegraphics[width=70mm]{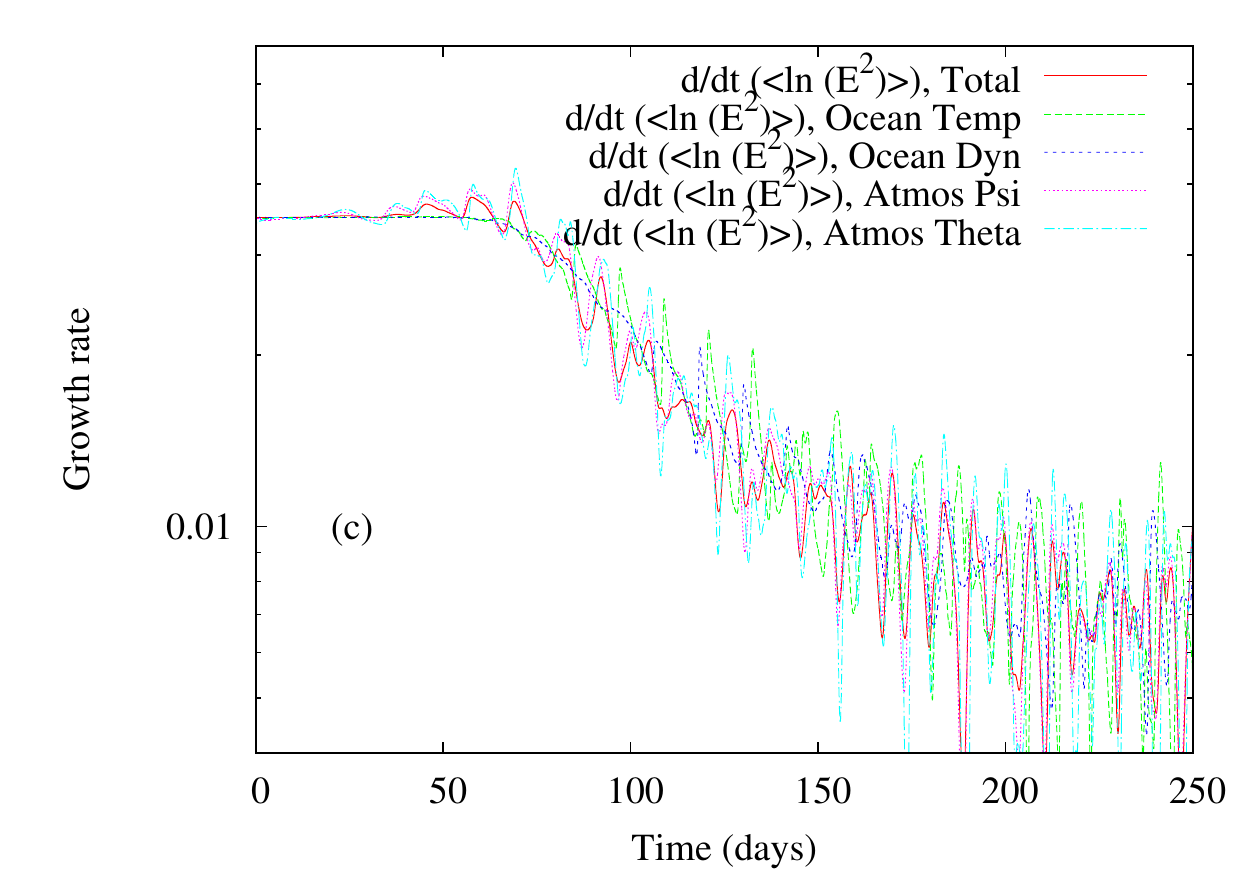}}
\caption{(a) as in panel (c) of Figure \ref{Fig:randerr} but with $d=6 \times 10^{-8}$ s$^{-1}$ ; (b) as in panel (b) of Fig. 7 but with $d=6 \times 10^{-8}$ s$^{-1}$; (c) as in 		
panel (c) of Fig. 7 but for  $d=6 \times 10^{-8}$ s$^{-1}$.}
\label{Fig:grate-d6x10-8}
\end{figure}

For the other considered value $d=6 \times 10^{-8}$ s$^{-1}$, the picture drawn above concerning the impact of the norm is even more pronounced, with a superexponential growth rate
of the $L^2$ norm reaching values of about an order of magnitude larger than the one associated with the largest LE (Fig. \ref{Fig:grate-d6x10-8}). 
Interestingly, the growth rate for the different variables at long lead times are close to each other, suggesting that the components of the system are all
developping on a similar time scale.  

\section{Concluding remarks}
\label{sec:concl}
In this paper we have explored the potential of the CLVs formalism to investigate the properties of a very relevant case of multiscale systems, namely coupled atmosphere-ocean models. The atmosphere and the ocean feature rather different time scales and are coupled through dynamical and thermodynamical processes. The understanding of the intimate properties of such a coupling bears great relevance in terms of basic climate dynamics, and, at practical level, for advancing our understanding of the predictability of the atmospheric and oceanic fields. Recently, coupled data assimilation is entering an operation stage, driven by the idea that understanding how errors propagate between the atmospheric and oceanic fields  is key to extending the predictability and achieving the dream of so-called seamless prediction \cite{Palmer2008}.  

Indeed, the climate model we study here is very simple, and is constructed through a severe truncation of quasi-geostrophic equations describing the dynamics of the atmosphere and of the ocean, provided with a simple yet a meaningful representation of their dynamical and thermodynamical coupling. Nonetheless, our model is well suited for exploring the mathematical and physical properties of interest here, and of potentially great relevance for more comprehensive models.

{\svc
We can separate the CLVs of the model in three main groups: \textit{a}, the unstable space, corresponding to the positive LEs (CVLs 1 and 2); \textit{b}, the stable space, corresponding to the negative LEs (CLVs 21-36); and \textit{c} the center direction, corresponding to the near-zero LEs (CLVs 3-20). The CLVs belonging to the group \textit{c}) are quasi-degenerate, as extremely often the angles between them are very small, and the corresponding FTLEs have an extremely high time correlation. The other CLVs have a non-pathological behaviour and the occurrence of quasi-tangencies is extremely rare. 
}

The variance of all CLVs is mostly distributed on the atmospheric variables, but notable differences emerge when looking at the role of the oceanic variables. For CLVs 1 and 2 we find that significant portion of the variance projects on the thermodynamic oceanic variables, while the contribution from the oceanic dynamic variables is basically nihil. This implies that the unstable modes are indeed coupled between the atmosphere and the ocean, with heat exchange being the dominant mechanism of coupling. Looking at CLVs of group \textit{b}, we find that the projection on the oceanic variables is almost vanishing, further reinforcing the idea that we are talking of quickly damped atmospheric modes. Important differences emerge within group \textit{c}: the variance for all CLVs 3-11 projects on all variables of the system, so that the slowest time scale of the system dominate, while for CLVs 12-20 the dynamic oceanic variables are entirely negligible, in agreement with the fact that in absolute value LEs 3-11 are much smaller than LEs 12-20 (yet also very small).

In the case of unstable and stable spaces, the corresponding FTCLEs obey accurately large deviations laws, which provide a solid basis for assessing the statistical properties of predictability of long to ultra-long time scales. Moreover, the large deviations laws obtained for the FTCLEs of group \textit{a} and \textit{b} correspond to those derived considering the FTFLEs and FTBLEs, whereas the statistical properties of the FTLEs on short time scales are rather different for the three geometrical constructions. Instead, such agreement of the large deviations laws is not found when looking at the FTLEs 3-20, which correspond to the degenerate CLVs. Note that if we construct large deviations laws for the FTLEs of the quasi-geostrophic model studied in \cite{Schubert2015a}, where no accumulation of LEs near zero is found as a result of the lack of ultralong time scales, we find in all cases almost perfect agreement between the exponents constructed using covariant, forward, and backward vectors (unpublished).  

This clarifies that the presence of geometrical degeneracies makes the understanding of predictability at long time scales more difficult. The basic reason for this is that, because of the multiscale nature of the system, it is difficult to disentangle the modes growing or decaying over very long time scales. Note that this point of view mirrors exactly the more intuitive idea, of direct relevance for seasonal to decadal predictions in the climate system, that if a system possesses both slow scales and fast scales of motions, our ability to perform accurate prediction over intermediate to long scales depends critically on our ability to define precisely the initial conditions for the variables (typically, oceanic ones) responsible for the slow time scales. 

Taking the CLVs point of view, in the system analyzed, it is hard for us to assess whether there is a true loss of nonuniform hyperbolicity, because many LEs are indistinguishable from zero, or, instead, the corrrect value of the LEs cannot be resolved unless one goes to extremely long time scales of observations. In fact, our model seems to qualitatively fit better the framework of partial hyperbolic systems \cite{Hasselblatt2005}, which allow for the presence of the so-called center directions, where the dynamics in neither really expanding nor contracting, and the decay of correlations can be non-trivial. Nonetheless, not even such a point of view is entirely satisfactory because partial hyperbolicity requires that in the stable and unstable directions uniform hyperbolicity is found, which is not the case here, as signaled by the fact that, \textit{e.g.} the rate functions of the positive FTLEs and of some of the negative FTLEs cross zero. Instead, a suitable mathematical framework is given by nonuniform partial hyperbolic systems \cite{Barreira2007}, where, basically, center directions corresponding to LEs indistinguishable from zero separate in terms of asymptotic behaviour the directions featuring asymptotic expansions (positive LEs) from those featuring asymptotic contraction (negative LEs). 

These considerations seem useful for briefly discussing the relationship between mathematical models of dynamical systems and actual physical systems with many degrees of freedom. The \textit{chaotic hypothesis} \cite{Gallavotti1995} says that chaotic systems of high dimensionality can be treated effectively as if they were Axiom A \cite{Ruelle1989}. Such an assumption is needed in order to construct a useful physical measure for the system (effectively, an {\sv Sina\"i-Ruelle-Bowen one}) and justify the application of \textit{e.g.} response theory \cite{Ruelle2009} for studying the impact of perturbation in high-dimensional chaotic forced and dissipative systems \cite{Lucarini2011,Lucarini2014}. 

In the presence of systems featuring strong multiscale properties, the chaotic hypothesis implicitly requires extremely long observation periods, in order to be able to potentially distinguish the various (long) timescales associated the small LEs. One may then propose a \textit{modified chaotic hypothesis}, by saying that chaotic systems of high dimensionality featuring multiscale behaviour can be treated effectively as partially hyperbolic systems, whereby we accept loss of information on the details on the dynamics of the ultralong time scales. In some cases, such systems have been shown to be able to support transitive properties and a  notion of physically-meaningful invariant measure \cite{Barreira2007} and to allow for the construction of a satisfactory response theory \cite{Dolgopyat2004} (see on the extension of response theory beynd the Axiom A case in \textit{e.g.} \cite{Baladi2012}). 

Note that time scale separation is not necessarily related to lack of hyperbolicity: the mathematical literature presents models of hyperbolic systems possessing different scales of motion and resulting from the perturbation of fast hyperbolic dynamics, namely the so-called solenoidal systems introduced by Smale \cite{Smale67} and  Wiliams \cite{Williams74}. 

The analysis of our model confirms that such conceptual difficulties reflect in the fact that  the dynamics of error in multiscale systems such as coupled atmosphere/ocean models does not fully conform to the standard point of view presented in  \cite{Kalnay2003} and commonly used for interpreting the predictability of atmospheric flows and the set-up of assimilation schemes. The averaged dynamics of the error along the CLVs has also been explored in the perspective of \cite{Nicolis1995}, and different behaviors were found depending on the specific norm chosen to 
measure the amplitude of the error. For the $L^2$ norm, a superexponential behavior is found, inducing a mean error amplification
in the stable subspace described by the CLVs 3-20. This behavior disappears when the logarithmic norm is used, except
for a few CLVs in the highly degenerate subspace from CLVs 6-10 for which complicate mixing and amplifications arise.  

Improving the understanding of predictability across time scales and across the atmospheric/oceanic domains, and our ability to construct coupled assimilation systems requires a deeper analysis of the complex dynamical processes hinted at in the present study, and in particular at an accurate analysis of the dynamics in the center directions, where strong geometrical degeneracy and coupling between the different CLVs takes place. The relevance of our results for motivating coupled data assimilations can be seen also from the fact that since the two CLVs spanning the unstable space project substantially also on the oceanic (thermodynamic) variables, the modern methods of data assimilation in the unstable space introduced by Trevisan \textit{et al.} \cite{Trevisan2004,Carrassi2008,Trevisan2010} need to take into consideration the ocean component of the system.

\subsection*{Acknowledgements}
The authors wish to acknowledge fruitful interactions with J. Baehr, V. Baladi, T. Bodai, A. Carrassi, G. Gallavotti, M. Ghil, T. Kuna, J. M. L\'opez, Y. Pesin, D. Ruelle, and S. Schubert. V. L. wishes to acknowledge the support of the DFG cluster of excellence CliSAP and of the FP7-ERC StG NAMASTE - Thermodynamics of the climate system (Grant No. 257106).  The work of S. V. is partly supported by the Belgian Federal Science Policy under contract BR/12/A2/STOCHCLIM.  Both authors acknowledge the support provided by the CliMathNet network.
This paper is dedicated to the memory of our friend and colleague Anna Trevisan, who recently passed away.

\end{document}